\documentclass[manuscript,screen]{acmart}
\usepackage{multirow}
\AtBeginDocument{%
  }

\setcopyright{acmcopyright}
\copyrightyear{2023}
\acmYear{2023}
\acmDOI{XXXXXXX.XXXXXXX}

\acmPrice{15.00}
\acmISBN{978-1-4503-XXXX-X/18/06}

\begin{document}

\title{Lessons Learnt from a Multimodal Learning Analytics Deployment In-the-wild}

\author{Roberto Martinez-Maldonado}
\authornote{Both authors contributed equally to this research.}
\email{roberto.martinez-maldonado@monash.edu}
\author{Vanessa Echeverria}
\authornotemark[1]
\email{vanessa.echeverria@monash.edu}
\affiliation{%
  \institution{Monash University}
  \streetaddress{20 Research Way}
  \city{Melbourne}
  \state{VIC}
  \country{Australia}
  \postcode{3800}
}
\affiliation{%
  \institution{Escuela Superior Politécnica del Litoral}
  \streetaddress{30.5 Via Perimetral}
  \city{Guayaquil}
  \country{Ecuador}
}

\author{Gloria Fernandez-Nieto}
\affiliation{%
  \institution{Monash University}
  \city{Melbourne}
  \country{Australia}
}

\author{Lixiang Yan}
\affiliation{%
  \institution{Monash University}
  \city{Melbourne}
  \country{Australia}
}

\author{Linxuan Zhao}
\affiliation{%
  \institution{Monash University}
  \city{Melbourne}
  \country{Australia}
}

\author{Riordan Alfredo}
\affiliation{%
  \institution{Monash University}
  \city{Melbourne}
  \country{Australia}
}

\author{Xinyu Li}
\affiliation{%
  \institution{Monash University}
  \city{Melbourne}
  \country{Australia}
}

\author{Samantha Dix}
\affiliation{%
  \institution{Monash University}
  \city{Melbourne}
  \country{Australia}
}

\author{Hollie Jaggard}
\affiliation{%
  \institution{Monash University}
  \city{Melbourne}
  \country{Australia}
}

\author{Rosie Wotherspoon}
\affiliation{%
  \institution{Monash University}
  \city{Melbourne}
  \country{Australia}
}

\author{Abra Osborne}
\affiliation{%
  \institution{Monash University}
  \city{Melbourne}
  \country{Australia}
}

\author{Dragan Gašević}
\affiliation{%
  \institution{Monash University}
  \city{Melbourne}
  \country{Australia}
}

\author{Simon Buckingham Shum}
\affiliation{%
  \institution{University of Technology Sydney}
    \city{Sydney}
  \country{Australia}
}

\renewcommand{\shortauthors}{Martinez-Maldonado et al.}

\begin{abstract}
Multimodal Learning Analytics (MMLA) innovations make use of rapidly evolving sensing and artificial intelligence algorithms to collect rich data about learning activities that unfold in physical learning spaces. The analysis of these data is opening exciting new avenues for both studying and supporting learning. Yet, practical and logistical challenges commonly appear while deploying MMLA innovations "in-the-wild". These can span from technical issues related to enhancing the learning space with sensing capabilities, to the increased complexity of teachers' tasks and informed consent. These practicalities have been rarely discussed. This paper addresses this gap by presenting a set of lessons learnt from a 2-year human-centred MMLA in-the-wild study conducted with 399 students and 17 educators. The lessons learnt were synthesised into topics related to i) technological/physical aspects of the deployment; ii) multimodal data and interfaces; iii) the design process; iv) participation, ethics and privacy; and v) the sustainability of the deployment. 
\end{abstract}




\received{20 February 2007}
\received[revised]{12 March 2009}
\received[accepted]{5 June 2009}

\maketitle

\section{Introduction}
Multimodal Learning Analytics (MMLA) is a relatively new area that formally emerged as such in 2012 at the ACM International Conference of Multimodal Interaction \citep{scherer20121st}. Since then, MMLA innovations have been opening exciting new avenues for supporting and generating a deep understanding of 
human learning by embracing the complexity of learners and their learning activities. MMLA systems are pushing the boundaries in educational research by de-emphasising the computational analysis of student interactions via online learning systems based on input devices such as the mouse and the keyboard \citep{worsley2021new}. In contrast, MMLA emphasises the many ways in which students interact with other students -- fully mediated \citep{vrzakova2020focused}, partly mediated \citep{schneider2018leveraging} or unmediated \citep{sumer2021multimodal} by technology -- with both teachers \citep{dmello2015multimodal}, and with physical learning environments \citep{yan2021footprints}. MMLA research also investigates learner attributes that are hard to automatically analyse without the use of specialised sensing systems, such as student emotions \citep{gorson2022using}, cognitive states \citep{mangaroska2022exploring}, distraction \citep{LIAO2022104599} and stress \citep{ronda2021towards}. In practice, MMLA endeavours often make use of sensors, such as eye-trackers, positioning systems, wearable microphones and physiological writs/chest bands, and advanced audio and video processing algorithms \citep{yan2022scalability,Alwahaby2022} that generate large amounts of multimodal data that can be analysed to gain a more holistic view of intrinsically complex human learning phenomena such as problem-based learning \citep{ma2022detecting}, effective collocated collaboration \citep{spikol2017estimation, Schneider16} and teamwork \citep{fernandez2021can}, self-regulated learning  \citep{azevedo2019analyzing}, engagement in adaptive online learning \citep{papamitsiou2020utilizing}, socially-shared regulated learning \citep{noroozi2019multimodal}, effective public speech \citep{ochoa2020controlled}, motor learning \citep{di2021keep}, and classroom teaching \citep{prieto2018multimodal,Ahuja2019EDU}. In a sense, MMLA aims at crystallising Mark Weiser's vision \citep{weiser1991computer} in educational contexts by creating learning spaces enhanced with ubiquitous computing capabilities to augment teachers' and students' activities. This can be achieved by creating applications that automate the capture of the lived experiences occuring in the learning space to allow later  access to those expierences by educational stakeholders \citep{Abowd2000} for the purpose of supporting reflection and learning \citep{Brotherton04}. Yet, enhancing learning environments with such sensing capabilities can easily impose practical challenges such as elevating the cost of implementation and making it harder to scale up the MMLA deployments in relation to fully digital learning analytics (LA) solutions.

Although recent literature reviews suggest that most MMLA solutions rely on video and audio analysis \citep{yan2022scalability, Alwahaby2022} (which should not be, in principle, too costly), MMLA researchers consistently report that most current MMLA systems only reach the prototype stage \citep{ochoa_multimodal_2022,cukurova2020promise}. Several practical challenges have been identified as potential threats to the broader adoption of MMLA that can ultimately impact the effectiveness of such innovations in supporting teaching and learning. Challenges include those related to the complexity of some sensor installation requirements and lack of technology readiness to enable full-scale deployments \citep{ochoa_multimodal_2022,cornide2019introducing}; lack of maturity from an analysis perspective to find causal relationships in multimodal data that can translate into actual improvements in learning outcomes \citep{Alwahaby2022}; misalignment between the technological MMLA innovation and the learning design \citep{sharma2020multimodal}; and ethical concerns that can ultimately dissuade educational stakeholders from adopting such complex solutions \citep{crescenzi2020multimodal,Alwahaby2022, cukurova2020promise}. While the number of small-scale laboratory studies -- conducted under controlled conditions -- is similar to the number of small-scale ecological (authentic) studies that have deployed sensors at the classroom level \citep{chua2019technologies}, the majority of MMLA studies (more than two thirds according to a recent literature review by \citet{yan2022scalability}) do not report enough methodological details to allow other researchers to replicate or learn from such deployments. Moreover, there is a lack of relatively large-scale studies conducted under authentic conditions that close the "data loop", that is, transitioning from multimodal data collection and analysis to the provision of some form of end-user interface \cite{yan2022scalability}, which is  one of the ultimate goals in LA research \cite{wise2021makes}. This makes it hard for MMLA researchers to fully understand the extent of the challenges involved in deploying MMLA innovations \textit{"in-the-wild" }(i.e., a deployment that is as naturalistic as possible). In-the-wild studies are commonly conducted in human-computer interaction (HCI) research with the aim of investigating the implications of embedding new technology interventions in everyday situations \citep{rogers2017research,crabtree2013introduction}. Conducting MMLA in-the-wild studies can thus contribute to understanding the challenges that need to be addressed to maximise adoption of MMLA innovations.

\begin{figure}[t]
  \centering
  \includegraphics[width=1\linewidth]{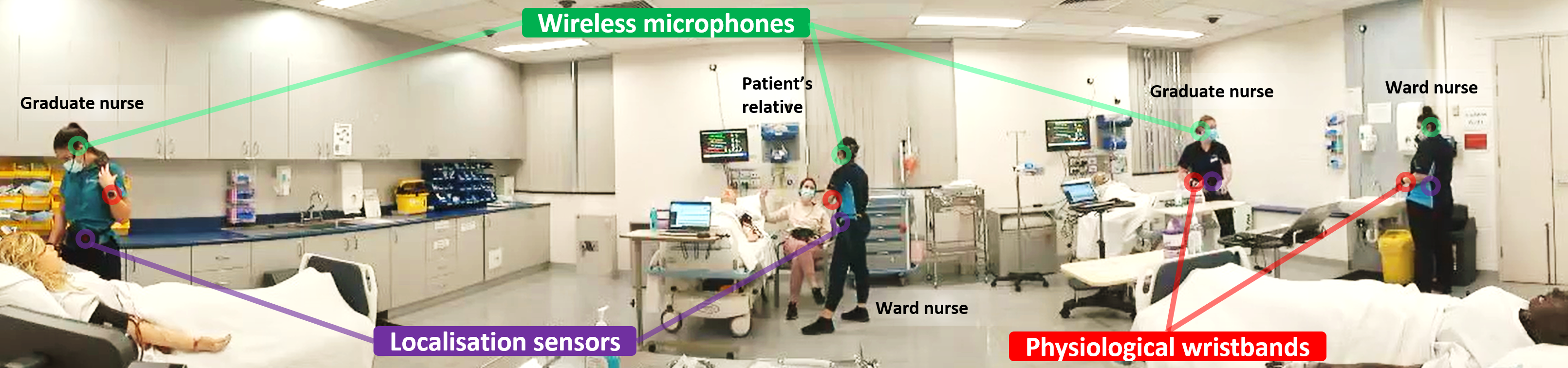}
  \caption{Sensors deployed in the high-fidelity simulation with a team of four students and two educators enacting the roles of a patient's relative and a doctor who provides some information.}
  \label{fig:space}
  \Description{}
\end{figure}

This paper addresses the above knowledge gap by synthesising a set of lessons learnt from a large human-centred MMLA study conducted in-the-wild (e.g., see Figure \ref{fig:space}). To the best of our knowledge, and according to the most recent MMLA systematic literature review \cite{yan2022scalability}, this is the first large, complex MMLA study that closes the LA loop by providing direct feedback to students on a group task using MMLA-based visual interfaces. A total of 399 consenting undergraduate students and 17 teachers participated in the study as a part of their regular classes. A range of sensing devices to capture multimodal data was used, namely, video recorders, microphones, physiological sensors, positioning sensors and an educator's logging system. The study spanned two years in which two deployment iterations were conducted. The first iteration focused on collecting multimodal data for research and design purposes. The second iteration focused on the deployment of a MMLA dashboard interface to support teacher-led, team reflection in the classroom. Evidence about the deployment included a set of interviews, surveys and focus groups with teachers, students and researchers to understand more about the practical challenges that emerged. 

\section{Related Work}

In this section, we review the most recent systematic literature reviews on MMLA and related works that have identified several prominent logistical, privacy and ethical challenges that need to be addressed for this promising area to remain relevant and have an actual impact on educational practices.

\subsection{Logistical Challenges}
Most MMLA studies so far have primarily focused on developing prototypes and testing the functionality of different combinations of sensors and analytics approaches \citep{shankar2018review, mu2020multimodal, noroozi2020multimodal}. Yet, many concerns have been raised regarding the logistical challenges that can emerge when moving from controlled settings to in-the-wild MMLA deployments such as the added intrusiveness of sensing devices and complexity in their installation and orchestration \citep{chua2019technologies}. \citet{yan2022scalability} systematically reviewed these logistical issues and identified a relatively low level of technology readiness regarding existing MMLA innovations, resulting in heavy reliance on the onsite support of researchers or technicians. This  undermines the sustainability of these systems and unnecessarily increases the complexity of the learning situation from the teachers' perspective. While most of the sensing technologies used in MMLA research can be purchased off-the-shelf, implementing these technologies in authentic physical learning spaces often requires extensive technical background for tasks such as physical installation, system integration, and modalities synchronisation \citep{crescenzi2020multimodal, shankar2018review, mu2020multimodal}. There is also a trade-off between data quality and affordability as most of the MMLA innovations that rely on mature sensing technologies, such as location sensors, eye-trackers, and biometric sensors, can be financially unscalable due to the high unit prices \citep{yan2022scalability}. Although low-cost alternatives are emerging \citep[e.g.,][]{ochoa2018rap, saquib2018sensei}, these technologies remain in the prototype and validation stages and often sacrifice accuracy or portability for affordability. 

Likewise, the lack of MMLA studies that have closed the LA loop by providing some form of end-user interface to students or insights to teachers make it harder for educational stakeholders to weigh the benefits against the potential added complexity to their already rich educational ecologies \citep{yan2022scalability}. Although the alignment between MMLA innovations and learning design should be one of the foundations for developing MMLA innovations \citep{cukurova2020promise, ochoa_multimodal_2022}, such alignments are rarely considered or reported in the existing literature, as noted in recent literature reviews \citep{sharma2020multimodal, praharaj2021literature}. This can undermine teacher and student confidence, if they do not understand how the MMLA system aligns with their teaching practices or learning outcomes. 

All of these challenges are hallmarks of emerging HCI infrastructures that must be co-evolved with work practices. This in-the-wild MMLA deployment offered the opportunity to study how both educational and technical stakeholders learnt to work together to address the challenges.

\subsection{Privacy Challenges}
As a research area that benefits from the data collection opportunities enabled by various sensing technologies, the privacy issues surrounding the adoption of MMLA innovations are the focus of critical debate. \citet{crescenzi2020multimodal} emphasised the need to consider the privacy implications of using sensing technologies to generate analytics about children's activity. Such implications have also been identified by students and teachers who have expressed concerns regarding the security of their data \citep{mangaroska2021challenges, kasepalu2021teachers}. These privacy implications of MMLA innovations have been under-investigated in the literature \citep{Alwahaby2022, yan2022scalability, Oviatt2018challenges}. Specifically, while most works published in MMLA  mention that informed consent was obtained from participants, none of the existing works has elaborated on the consenting strategies they adopted, which could contribute valuable insights regarding data security measures for protecting individual privacy and maximising data autonomy (e.g., individuals' autonomy of removing their data from the database) \citep{beardsley2020enhancing}. Additionally, while most of MMLA innovations endeavour to provide dashboards and visualisations for supporting educational practices, privacy issues regarding who has the right to see these visualisations  remain unclear, especially in the contexts of collaborative learning where, in most cases, individuals' personal trace data, even anonymised (e.g., masking students' identity with numbers or colours), could remain identifiable when used for provoking reflections at a group-level, since other students typically have the contextual knowledge to decode anonymised representations \citep{mangaroska2021challenges, Alwahaby2022}. Providing additional empirical evidence on educational stakeholders' perspectives of these privacy-related issues could potential benefit the on-going development of MMLA, and is a particular focus of this study.

\subsection{Ethical Challenges}
Beyond logistical and privcy issues, the potential biases in analytics, and cognitive dissonances that may be caused by the inconsistency between individuals' observations and generated insights, could also undermine the potential benefits of MMLA innovations \citep{ferguson2016guest,Oviatt2018challenges}. Such issues are vital as the accuracy of the existing MMLA-based predictive models and early-warning systems are far from suitable for practical deployment (e.g., rarely above 80\% accuracy), and these models have mostly been developed and evaluated based on relatively small sample sizes (i.e., with n < 50) \citep{yan2022scalability}. These small sample sizes combined with the poor reporting standards found in the existing MMLA literature could also mask potential algorithmic biases that may disadvantage certain minority groups of students as replicating these studies remain difficult without adequately reported methodologies \citep{luzardo2014estimation, yan2022scalability}. Additionally, \citet{Alwahaby2022} also highlighted the significant concerns regarding the need to enhance trust and data transparency within MMLA systems and \citet{yan2021footprints} suggested that more research needs to be done to assess the potential risk of making decisions with incomplete multimodal data.
Consequently, understanding the ethical practices of using these analytics is also essential but rarely considered in prior literature \citep{selwyn2019s} and requires the participation of key educational stakeholders such as students and educators \citep{Oviatt2018challenges}.  A large-scale in-the-wild study opens new opportunities to study approaches to these ethical challenges under more authentic conditions than has been reported to date. 

\subsection{Contribution to HCI and Research Question}
Against the literature reviewed above we formulate the following research question (RQ) that guided our study: 

\textit{\textbf{RQ:} What logistical, privacy and ethical challenges emerge from a complex MMLA, in-the-wild study that closes the analytics loop by providing direct feedback to students?}

In addressing this question, the contribution of this paper is a set of lessons learnt regarding how such challenges were, or could have been, addressed in the context of a two-year deployment of a MMLA system in an authentic educational scenario. The implications of this study should assist researchers, developers and designers in making informed decisions about the effective deployment of innovations that involve the use of ubiquitous computing technologies, sensing devices and artificial intelligence (AI) algorithms to augment teaching and learning in physical spaces. 



\section{Study in the wild}

\subsection{Context}
The study presented in this paper followed a human-centred learning analytics approach \citep{BuckinghamShum2019}. A partnership between a team of four teachers (\textit{senior teaching team}) and four LA researchers (\textit{researcher team}) was forged to progressively co-create a MMLA innovation to be embedded into the regular classes of an undergraduate course of the Bachelor of Nursing program at Monash University. Students were also consulted to understand i) the extent to which their lived learning experiences can be impacted by the use of sensing technologies, ii) their ethical and privacy concerns, and iii) the extent to which the data insights can effectively support their learning. While details about the co-creation process go beyond the scope of this paper, and can be found elsewhere \cite{Huceta22,fernandez2022beyond}, key information about educators' and students' involvement is provided below. 

In the targeted course, high-fidelity, immersive team simulations are typically conducted to help students develop effective collaboration and communication skills while learning from errors in a safe environment \citep{Sarcevic2012}. High-fidelity simulation is a healthcare education methodology, conducted in a realistic but simulated health setting environment, where clinical situations that students may encounter in the workplace are reproduced using sophisticated manikins as patients \citep{maran2003low}. In these simulations, students are often posed with a situation that they need to address without the instruction of a teacher, followed by a reflective \textit{debrief}, facilitated by a senior teacher, in which students reflect on their actions and learning. The educational goal of the MMLA deployment was decided by the senior teaching team, and aimed at improving the provision of feedback to students in the debrief.

\subsection{Study iterations and participants}
The MMLA study had two iterations. The first, conducted in 2021, focused on i) collecting a rich multimodal dataset, ii) enhancing the understanding of the senior teaching team about the possibilities enabled by the multimodal data, and iii) asking students about envisaged uses of their data and potential concerns regarding the use of sensors in their regular learning spaces. The second, conducted in 2022, focused on i) closing the LA loop by deploying MMLA visual interfaces to support the reflective debrief, and ii) expanding the multimodal dataset. Both iterations were conducted under almost identical conditions: the same course, learning goals, senior teaching team, and lesson design.

A total of 399 students consented to participate in the study (261 -- 196 females -- consenting out of 461 enrolled students in iteration 1; and 138 -- 114 females -- out 358 in iteration 2). Some students were invited to follow-up activities for them to provide their feedback based on their lived experiences (see details in the next section). Besides the four senior teachers, another 13 teachers were involved in both years facilitating the lesson plan of the simulations.


\subsection{The authentic learning situation}
Each 3-hour class was typically attended by 10-15 students and was conducted across two learning spaces: a regular classroom and the specialised simulation classroom. The latter featured four beds with a patient manikin in each of them as shown in Figure \ref{fig:space}. Two consecutive simulations would be conducted during the class, both focused on prioritising care and identifying the deteriorating patient who required urgent attention. Students were given important information called a \textit{handover} before commencing, and then asked to provide care in teams according to the assessment they conducted on each manikin.  Each team was made up of four students who volunteered to play either a graduate or ward nurse. These students were also asked to optionally consent to be part of the study. Two teachers enacted the roles of a patient’s family member and a doctor assisting with patient care after students identified a problem.  Other students were invited to be observers, watching the simulation unfold. Immediately after each simulation, a whole class debrief conversation was led by a teacher in the regular classroom.



\begin{figure}[t]
   \centering
   \includegraphics[width=\textwidth]{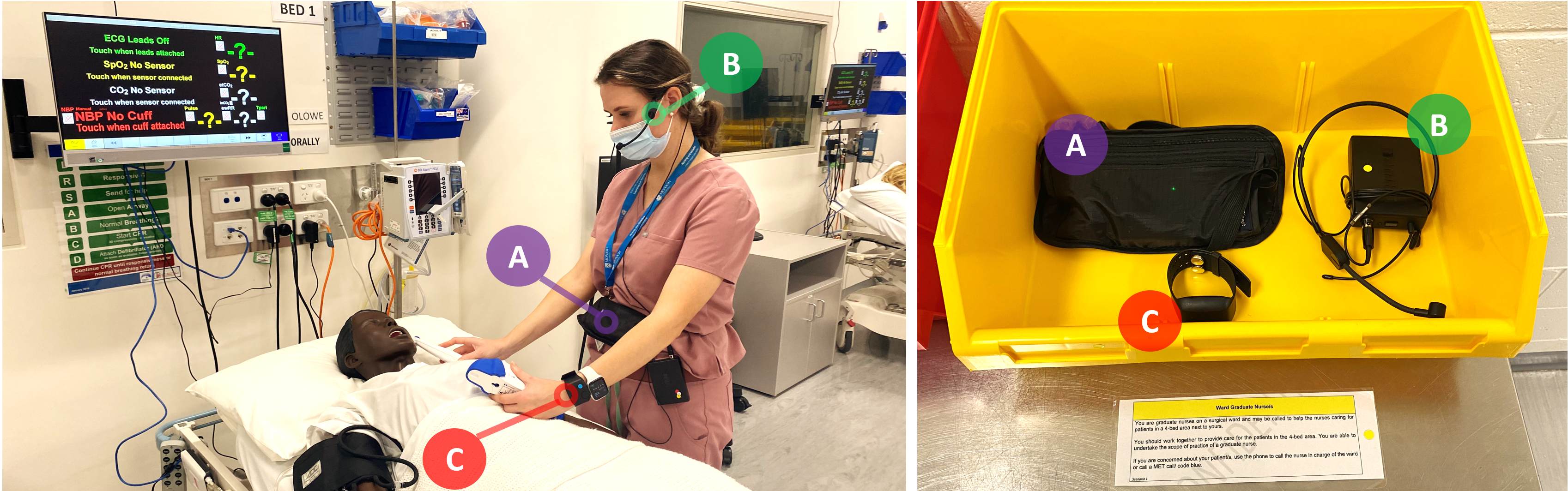}
   \caption{Multimodal Learning Analytics (MMLA) deployed in an authentic healthcare education setting. Left: Illustration of a student wearing the sensors during a team simulation. Right: Each sensor set to be worn by each student within a team (A--an indoor positioning locator inside a belly bag, B--a wireless microphone and C--a physiological wristband) was placed by the teaching team on coloured trays to enable easy access to and organisation of the sensors during and between classes.} 
   \label{fig:equipment}
 \end{figure}

\subsection{Apparatus}
Before entering the simulation classroom, consenting students were asked to wear a number of devices, namely, a wireless headset with an unidirectional microphone; a physiological Empatica E4 wristband, with built-in sensors that capture heart rate variability, electrodermal skin activity among other physiological measures; and a Pozyx indoor positioning locator, with built-in sensors that capture \textit{x-y} position and body orientation of each student in the learning space (see Figure \ref{fig:equipment}). Each set of sensors was colour-coded according to the role enacted by each student (i.e., red and blue for graduate nurses, and yellow and green for ward nurses). The simulation room was already equipped with a set of built-in video cameras, and an additional 180-degree video camera was added to the set up. All the data (except those captured by the wristbands due to limitations of the software vendor, to be discussed later in the lessons learnt) were captured and synchronised in \textit{real-time} using our open-source MMLA infrastructure called \href{https://teamwork-analytics.github.io/yarn-sense}{YarnSense}.

It should be noted that despite these devices, this experience was not completely novel to students, since these simulations are commonly video recorded and students often wear lapel microphones for the rest of the class to observe the simulation from the regular classroom. After each simulation, the reflective debrief was conducted with both the team who participated in the simulation and the rest of the class in the regular classroom. A dashboard displaying representations of the multimodal data collected during the simulation was deployed in the debriefs as detailed next.

\subsection{The MMLA dashboard}
\begin{figure}[h]
   \centering
   \includegraphics[width=\textwidth]{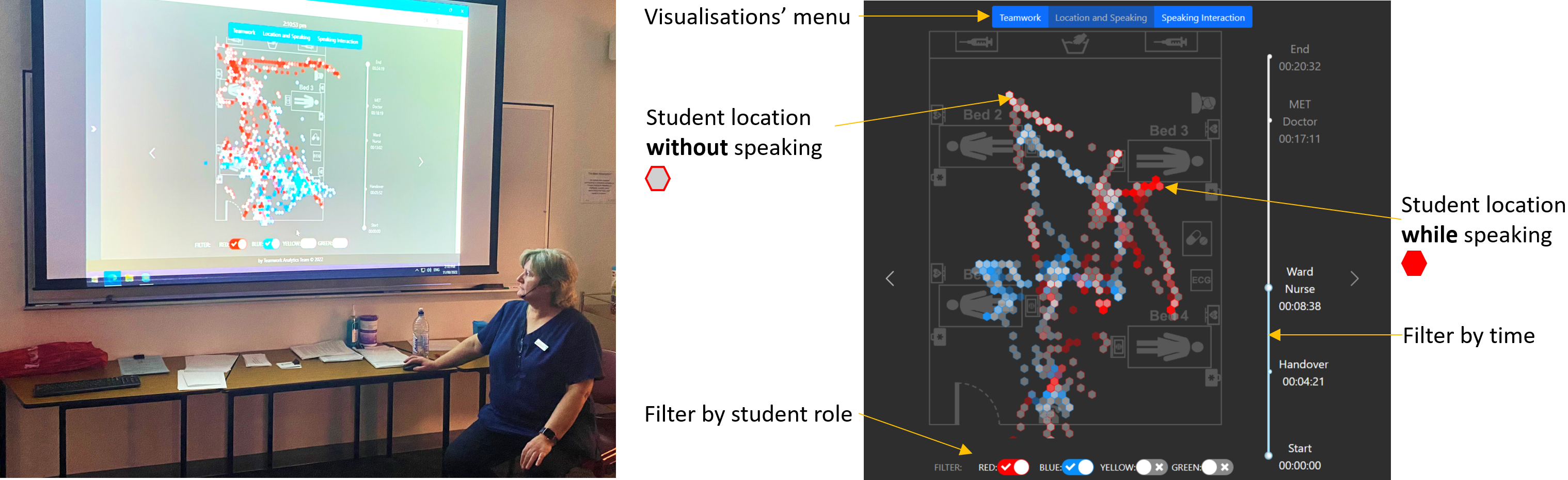}
   \caption{Left: A teacher leading a team debrief using positioning and audio data in the MMLA dashboard. Right: The MMLA dashboard providing a menu of visualisations at the top. The selected example visualisation shows whether students, according to their role/colour, were speaking or not at certain positions of the learning space (see hexagons filled with student's colour or in grey, respectively).}
   \label{fig:dashboard}
 \end{figure}

Figure \ref{fig:dashboard} depicts the MMLA dashboard deployed in the second iteration of the in-the-wild study. This interface was automatically loaded after each simulation and shown in the main screen of the regular classroom (Figure \ref{fig:dashboard} left), to be used by teachers to support the reflective debrief. A set of three MMLA visualisations was designed with the senior teaching team, each using more than one modality of data. For example, Figure \ref{fig:dashboard} (right) illustrates the dashboard showing one of such visualisations. This shows the locations of each colour-coded student in the simulation space where they were speaking with each other or with the patient, by automatically triangulating coordinate data from the positioning sensors and outputs using a voice activity detection algorithm applied on the multi-channel microphone signals. Each hexagonal data point summarises which student was in that position doing most of the talking. 
Two other visualisations included a bar chart to summarise positioning data coded according to the particularities of the team task; and a sociogram (a network based chart) focused on depicting the extent of verbal communication among team members, the patient and the other actors. Further details about these go beyond the purpose of this paper which is focused on reporting the lessons learnt from the iterative deployment as a whole. Nonetheless, details and the source-code can also be found in our \href{https://teamwork-analytics.github.io/yarn-sense}{YarnSense} repository.

\section{Methods}

\subsection{Research "in-the-wild"}
The phrase "research in-the-wild" is used in HCI studies to differentiate between research conducted in lab-based environments and research that involves embedding new technology interventions in everyday situations \citep{crabtree2013introduction}. Arguably \citep{rooksby2013wild}, a key tenet of studies conducted in-the-wild is that they can provide more ecologically valid findings compared with typical measures collected under controlled conditions \citep{balestrini2020moving}. Indeed, in-the-wild studies often have to deal with a number of practical and ethical challenges and uncertainties rarely discussed in published studies \citep{rogers2017research}. Carefully identifying these can reveal the biases and the logistics that researchers and designers should consider to design and deploy emerging technologies in a specific context. Based on \citeauthor{rogers2017research}'s proposed framework \cite{rogers2017research} to design for and analyse in-the-wild research, other authors, such as \citet{balestrini2020moving}, have suggested analysis approaches to identify practical challenges that can arise in an in-the-wild study. Inspired by this work \citep{balestrini2020moving}, we explored the main logistic, privacy and ethical challenges when designing and deploying MMLA research in-the-wild, organised around five main themes: 1) \textit{space and place} -- what is the impact of the computational system on the existing setting?; 2) \textit{technology} -- how was the technology used? -- in this case, the data and the analytics; 3) \textit{design} -- what was the impact of the design approach? -- in this case, human-centred design; 4) \textit{social factors} -- what are the stakeholders' concerns and expectations?; and 5) \textit{sustainability} -- what is needed for the new technology to be used continuously over time? 

\subsection{Sources of evidence}
Table \ref{table:evidence} summarises the sources of evidence captured from students, teachers, and the researcher team, using a set of interviews and surveys. This section describes how the questions asked in these cover the five themes introduced above.

In the first iteration of the study, all participating students were asked to rate their perception of the intrusiveness of the sensing technology in the learning space (theme 1) using a seven-point Likert scale. In addition, twenty volunteering students (18 females, avg. age: 22.21, std. dev.: 4.40 - S1-S20) who participated in the first iteration also participated in 1-hour post-hoc individual interviews to explore their perceptions and experiences more in-depth  in relation to the themes \textit{1- space and place} (i.e., their perceptions on the intrusiveness of devices used and the data collection); \textit{2- data and analytics} (their perceptions on the MMLA dashboard); and \textit{4- social factors} (their concerns regarding the deployment). Moreover, students were presented with early prototypes of the visualisations that ended up in the MMLA dashboard in the second iteration using their own data. Each interview was recorded using an online video conferencing platform (i.e., Zoom) and lasted about 60 minutes.

\begin{table}[b]
\caption{Sources of evidence and themes explored inspired by \citeauthor{balestrini2020moving}'s approach \cite{balestrini2020moving} to identify practical challenges that can arise in an in-the-wild study}
\label{table:evidence}
\begin{tabular}{|l|l|l|l|}
\hline
\textbf{Iteration}       & \textbf{Participants} & \textbf{Sources of evidence} & \textbf{Themes explored}                                                                                                                           \\ \hline
\multirow{1}{*} & Students              & Survey (N=253)        & 1) Space and place                                                                                                                                 \\ \cline{2-4} 
                    & Students              & Interviews (N=20)            & \begin{tabular}[c]{@{}l@{}}1) Space and place\\ 2) Data and analytics\\ 4) Social factors\end{tabular}                                             \\ \hline
\multirow{2}{*} & Students              & Survey (N=47)         & 2) Data and analytics                                                                                                                              \\ \cline{2-4} 
& Teachers              & Survey (N=11)         & 2) Data and analytics                                                                                                                              \\ \cline{2-4}
                    & Teachers              & Interviews (N=4)             & \begin{tabular}[c]{@{}l@{}}1) Space and place\\ 2) Data and analytics\\ 3) Human-centredness\\ 4) Social factors\\ 5) Sustainability\end{tabular} \\ \cline{2-4} 
                    & Researchers           & Survey (N=5)          & \begin{tabular}[c]{@{}l@{}}1) Space and place\\ 3) Human-centredness\\ 4) Social factors\\ 5) Sustainability\end{tabular}                         \\ \hline
\end{tabular}
\end{table}

In the second iteration, the four senior teaching team members (T1-T4, all females) and a total of 7 additional supporting teachers (T5-11, also all females) led the debriefs after each simulation across all classes. All teachers who led the debrief using the MMLA dashboard were asked to complete a survey immediately after each class. The survey comprised questions related to the second theme of the study in-the-wild (\textit{technology - data and analytics}) to inquire about the helpfulness (e.g.\textit{ How the visualisations assisted you in the reflective debrief?}) and integration of the MMLA dashboard into the learning experience (\textit{e.g., How did you integrate the visualisations into the debrief?)}. Moreover, the four senior teaching team members, who were also part of the teachers leading the debrief (T1-T4), participated in a post-hoc reflective interview session to gain insights into their experiences and challenges in using the MMLA dashboard in an authentic setting. We asked questions related to the five themes presented above to explore their perceptions of the study in-the-wild (e.g., regarding \textit{1- space and place}: \textit{How intrusive was it to equip students, teaching staff and the simulation space with various sensors? What kind of unexpected issues did you face that may have affected the learning goals of the simulations?}; \textit{2- technology}: \textit{How and for what purpose did you use the tool during the debrief?}; \textit{3- design and human-centredness}: \textit{To what extent do you value the collaboration with researchers to design this technology with them?}; \textit{4- social factors}: \textit{Did you perceive or hear any concerns from students regarding the study?}; and \textit{5- sustainability}: \textit{What steps would be needed to make the current system into a real-world application without the help of researchers behind it?}). Two interviews were conducted with two teachers at a time, each lasting about 40 minutes. Video recordings of all the interviews were fully transcribed for further analysis. 

In addition, students who consented to participate in the second iteration were invited to participate in a survey to gather their perceptions on the trust of the MMLA dashboard information and their comments about their experience when navigating the information presented in the MMLA dashboard(2 - data and analytics). 
The survey showed the visualisations that were included in the dashboard using students' own data. Students were asked to rate their perception of trust using a five-point Likert scale (1= \textit{I would completely trust this information} - 5: \textit{I would not trust this information}). They were also asked to explain their rate and give comments on their experience with the MMLA dashboard. A total of 47 students (40 females, avg. age: 23.81, std. dev: 5.61 - S21-67) completed this survey.
 
Finally, five members of the research team (R1-R5), who were the ones mostly involved in the deployment, were asked to fill in a survey to reflect on and document their challenges and experiences during both the first and second data collection. The questions were related to four themes (except technology since it is about how teachers and students used the technology) and were similar to those asked to teachers as presented above. The complete protocols for teachers' post-reflection sessions and the research team survey can be found in the supplementary material (see appendices A.1-5, in the order the surveys and interviews were described above).

\subsection{Analysis}

We synthesised a set of lessons learnt by following a hybrid deductive and inductive thematic analysis approach \citep{fereday2006demonstrating, thomas2008methods}. The \textit{researcher team} and the \textit{senior teaching team} met several times to discuss challenges and concerns around the logistics and consenting strategies that may have impacted the learning experience of students and the overall teaching experience. The first deductive step involved using the literature on research in-the-wild \citep{rogers2017research, balestrini2020moving} to identify the initial five themes presented above that also helped to scaffold the collection of further evidence about the MMLA deployment. Then, the sources of evidence were inductively coded and triangulated by three researchers together to find emerging topics for each main theme \citep{braun2012thematic}. None of these researchers was involved in responding the research team survey. Decisions were made simultaneously and through consensus \citep{mcdonald2019reliability}. All the authors then met to further discuss the lessons learnt and how the evidence can illustrate the challenges that were faced, considering the perspectives of the various educational stakeholders. 

\section{Lessons learnt}

\subsection{Space and place}
This theme included the following four topics: i) intrusiveness of sensors; ii) technology readiness; iii) unexpected issues during the deployment; and iv) the trade-off among data quality, portability of sensors and affordability. 

\subsubsection{\textbf{Intrusiveness of sensors}}
This topic focuses on teachers' and students' perceptions of the intrusiveness of the MMLA deployment as a whole, and the wearables sensors in particular. 
All teachers agreed that the whole MMLA deployment was less intrusive than what they had initially expected. As expressed by T3: \textit{"when we look back at the previous year, we thought it was going to be a lot more intrusive than it turned out to be"}. Yet, two teachers also mentioned that the deployment may have caused \textbf{distraction to some students}, as expressed by T1: \textit{"students may be distracted by the research happening instead of being aware of the actual learning"}. For example, one observation made by the same teacher was that \textit{"sometimes researchers were entering and leaving the debriefing room to solve some technical issues"}.

Teachers had mixed reactions about the intrusiveness of the wearable sensors. Two teachers (T3 and T4) explained that the sensors were not intrusive to students while two other teachers (T1 and T2) believed that \textbf{sensors may have been uncomfortable for some students to wear}. T2 mentioned that \textit{"the belly bag was too tight for some students"} and that \textit{"the microphone fell off for some students"}. These issues may cause students to be \textit{"self-aware about the research"} (T2). 
In contrast, most of the 261 students who participated in the first iteration reported that they felt comfortable wearing the sensors during the simulation (Q: \textit{"I felt comfortable wearing the sensors during the simulation"}, Median = 6, mean = 6.166, std = 1.07, min = 1, max = 7). In the more in-depth interviews with students, most of them (N=18 out of 20) also expressed that it was fine to wear the sensors during the simulation (e.g., \textit{"It was fine wearing the sensors. I didn't mind wearing them"} - S9). Some students explained that, at the beginning, they felt \textit{conscious} (S20), \textit{nervous} (S9) or \textit{stressed} (S12) about being monitored. However, once they started the simulation, they focused more on their learning task than on the sensors. One student stated: \textit{"I was focused on my simulation. So, after I had them in and went into the simulation, it was fine. I didn't notice them."} (S17). Students also pointed out that it is normal for nurses to wear various devices or instruments. In this sense, wearing sensors was not different from their usual practice. One student explained this as follows: \textit{"In placements, we’re so used to wearing a pickpocket on the side. So it didn't matter to me because I was already used to that"} (S19). 

When students were asked if they had any \textbf{concerns related to health and safety} while wearing the sensors (since the first iteration of the study was conducted in between lockdowns, during the COVID pandemic), all the interviewed students (except one) expressed that they did not feel any concern about this. Only one student pointed out the following: \textit{"when I was putting the equipment on, I was wondering whether it was clean"} (S20). Students mentioned they knew that the University had strict safety protocols for running these simulations and that, in clinical environments, it is a regular practice to have all equipment clean after every use. 

In sum, although some teachers were somewhat concerned about the intrusiveness of adding wearables to the learning environment in terms of students' comfort and potential distraction, most students reported that they were comfortable with wearing the sensors, citing that they commonly wear various devices in simulation-based learning sessions and at their placements. Yet, it is critical to follow strict hygiene protocols, which was the case in both iterations.

\subsubsection{\textbf{Technology readiness}} 
This topic focuses on the potential impact of the lack of MMLA technology readiness. From a research perspective, two members of our research team (R1 and R2) explained that whilst the off-the-shelf sensors utilised in this setting were easy to install, there still is complexity in maintaining the ecology of devices during the deployment. Although wearable microphones are common in healthcare simulation, smart watches are already being worn by students to monitor themselves and indoor positioning technologies based on smart-phones already exist (typically less precise than the one used) \citep{gualda2021locate}, the \textbf{technology was not yet ready to enable bring-your-own-device (BYOD) strategies} to work smoothly in our MMLA deployments. Hence, we still needed to provide specific devices to students and they needed to be worn correctly to maximise the quality of the data collected. Moreover, all sensors were connected to a data infrastructure running on more than one computer, since different sensor vendors required particular specifications in terms of hardware and operating system configurations. This made it hard, if not impossible, to easily connect all sensors to a single machine. Consequently, a team with technical expertise was still required to closely monitor the data infrastructure and to help students to put the sensors on.

Teachers noted the \textbf{potential negative impact on the lesson plan} as a result of asking students to wear the sensors correctly during a regular class. T1 noted: \textit{"If we factor in more time for that [sensor equipping], we still only have a three-hour session, which means that we cut into time for other things. So we need to think about ways that we can reduce the amount of time spent equipping students."} However, T4 had a contrasting point of view, expressing that \textit{"there was no great time-saving difference between teams wearing the devices compared to those who did not"}, suggesting that this is not a time-consuming task. Nevertheless, from the experience in the first iteration, the senior teaching team devised ways to equip students with the sensors more efficiently. Thus, the lead course coordinator placed coloured trays (blue, yellow, green, and red) per each student's role to put the devices inside the tray (e.g., see Figure \ref{fig:equipment}, right) so students could easily equip themselves. This new configuration was well received by T1, T3 and T4, stating that it made it easier to identify the equipment and the roles (e.g. \textit{The coloured [trays] worked really well, I think that was great.} - T1; \textit{It was easier to equip students this year [second iteration] than last year [first iteration]}. - T3). 

During the second iteration, we also observed teachers instructing students on how to wear the devices properly. This illustrates how \textbf{teachers became involved in developing their own strategies to appropriate the MMLA sensing technology} and configure the space without the intervention of the researcher team. Teachers also proposed ideas to overcome the issues about technology readiness by suggesting that, in the future, clear instructions could be provided to students to wear the equipment correctly without any assistance, for example, using \textit{"big posters showing pictures of how the devices should be used properly"} (T1). In terms of readiness of the infrastructure, teachers were eager to use \textit{"on-the-go"} packed and ready-to-use sensors (T1 and T2) to adopt the tool in their learning environment without the researchers' intervention: \textit{"We can use the system without the need of the research team, we want to appropriate the use of it without needing assistance" (T2)}. This is linked to the fifth theme on the sustainability of the deployment further explained later.

\subsubsection{\textbf{Unexpected technical issues during the deployment}}
This topic focuses on unenvisaged problems with the sensing technologies deployed in the learning space that impacted data collection and the analytics. 
Our research team reported various \textbf{technological issues when configuring the devices}. For example, while setting up the space before the second iteration of the study, the research team found that a third-party driver for the positioning sensors had to be updated thus making the installation time longer than expected for the MMLA system to work exactly as in the first iteration. Other issues emerged during the study itself. R2, R3 and R5 reported that, although the physiological wristbands used in the study are marketed as high-end sensors and are relatively new, some stopped working unexpectedly (e.g., \textit{"Two [physiological wristbands] were not working or they turned off in the middle of some classes"}- R5). The research team had to investigate the cause of this malfunction and re-install the firmware in between class sessions with mixed results. Also during the study, the University-owned computers restarted unexpectedly due to software updates controlled centrally by the university's technical support team. R3 mentioned that: \textit{"the [positioning sensors] laptop restarted due to University device update"} and R5 complemented that due to this issue, the visualisations could not be generated: \textit{"one of these devices restarted on its own even in the middle of data collection, and all visualisations are depending on this device"}. The computers that were used during data collection are operated by the University, leaving the research team with limited freedom to configure them as the need arises.   

Another unexpected issue was that some \textbf{students wore the sensors incorrectly} causing data loss. As noticed by R5, some students did not wear the microphone properly (e.g., "placing the mic far from the mouth"), even though the instructions given to them were clear. This could have affected the audio quality for a few students: \textit{"The talking volume of students actually have great influence on audio data collection, if the volume is high, then the audio signals might be harmed, if the volume is low, then some audio might not be clearly recorded"} (R3). These unexpected issues hindered the correct generation of visualisations or yielded to an incomplete dataset collection for some teams.

Most of the issues described above are expected given the particular requirements imposed by multiple third-party technologies used in our MMLA deployment. While some of these technical challenges go beyond the control of the research team, researchers reflected and provided suggestions to minimise these in the future, such as replacing the current physiological wristbands with alternative wearables even though they may collect less granular data (i.e., a Fitbit sense) (R5) and reduce the number of computers used in the deployment to minimise connectivity or operating system related issues (R2 and R5):\textit{ "In total, we have 3 computers for data collection. It would be better to merge to two"} (R2).

\subsubsection{\textbf{Trade-off among data quality, portability of sensors and affordability}}
This topic focuses on how the sensing technology used can impact data quality, and the portability and affordability of the MMLA system.
A trade-off among these three aspects of the deployment can be illustrated in terms of audio data collection. Six pairs of inexpensive, minimalist wireless microphones (Xiaokoa 2.4G -- around USD\$17 each) were used in the first iteration with the aim of minimising intrusiveness caused by the wearables. These were consisted of one slim headset without any further equipment or antenna, thus, making them highly affordable and portable. However, the lack of a less portable signal transmitter in these devices made the signal inconsistent and subject to interference. This resulted in only 23 out of 57 simulation sessions being fully recorded for all team members. A second quality issue was the voice overlap, since other students' voices were easily captured when in close proximity. Without special audio post-processing, this led to inaccurate data analysis results. To address these issues, the microphones were upgraded in the second iteration. The new microphones (Shure BLX14P31 Wireless Headsets) resolved the first quality issue about signal interference, as they provided multiple frequency bands for signal transmission. The second issue about voice overlap was also resolved, as these microphones were unidirectional, minimising the chance of getting voice overlap. Although the new microphones resolved the quality issues, it made the MMLA system less affordable as the price of each pair is around USD\$300, costing around 20 times more than the previous microphone, and they included a bodypack transmitter that students had to wear on their hip additionally to the headpiece connected through a cable, making them slightly more complex to wear. Each microphone also required a receiver in the form of a black box connected to the audio infrastructure, making the setup more complex too.  

In sum, obtaining high-quality data in a MMLA deployment was prioritised, but required more expensive and bulkier sensing devices, a tradeoff that must be weighed against the expected instructional benefit. For example, better audio data quality can lead to the careful study of the quality of students' conversations with the aim of supporting learning effectively (e.g., \cite{zhao2023mets})

\subsection{Data and analytics} 
This theme included the following three topics: i) the purpose of capturing multimodal data; ii) data incompleteness and trustworthiness; and iii) emerging issues related to the MMLA Dashboard. 

\subsubsection{\textbf{Purpose of capturing multimodal data}}
This topic focuses on students' and teachers' perceptions of the educational purpose of the MMLA deployment. 
Regarding whether students understood  \textbf{what data was being collected}, all interviewed students (N=20) remembered they wore a bracelet (i.e., \emph{``a watch'', ``a little hand band'', ``a wristwatch''}), a microphone (i.e., ``a headphones'', ``a headset''), and an indoor positioning device  (i.e., \emph{``a fanny pack'', ``belly bags'', ``a tag at the front''}). Yet, only two students realised that video data were also recorded (\emph{i.e., ``you recorded us'' and ``there was video''}) possibly because video recording has been a staple in this educational context  \cite{megel2013high}.
In the first iteration, when students were asked about their perceptions regarding the purposes behind \textbf{why their data were collected}, a half of them (N=10) indicated that data was collected for others (researchers or teachers) to understand how students perform during the simulation scenario and assist future students in their learning (e.g., \emph{``[researchers] can suggest ways that the [teachers] can change to help or assist students in the simulation, the common things that students are struggling with''} (S17). Likewise, S4 noted that the data was collected to \emph{``help [future students] have better learning resources and using us as guinea pigs to see what needed to be improved for the learning outcomes for future nurses''}. Various students (N=11) concretely explained that the new technology was to support them directly. For example, students claimed that it was \emph{``to see how everyone reacts when there's a situation going on and how people prioritise and react''} (S7) and \emph{``to provide some feedback about the whole scenario to students and educators''} (S1). Other purposes that students mentioned were related to the assessment of the simulation effectiveness (N=4) (e.g., \emph{``to see the effectiveness of the simulation, or whether it's working and if it is beneficial to our learning''} S1), the support for future nursing professionals (N=4) (e.g., \emph{``to see what we experience and we make, for improvement of future nurses''} -- S10), and only two of them were not sure about the purpose of the MMLA deployment as a whole.

In relation to the second iteration, \textbf{teachers felt satisfied with the use of the debrief MMLA dashboard}. All senior teachers highlighted that the debriefing tool helped to \textit{reinforce and back up the discussion during the debrief}, stating that the tool \textit{"ignited some light bulb moments around this discussion"} (T4). As stated by T4, the tool can draw attention to key points and spark discussions towards improvement in their practice: \textit{"I used the data to draw attention to improving communication. I also used it to discuss the most useful aspects of teamwork and communication students needed to learn for a deteriorating patient"}. Specifically, the tool was used to validate and emphasise some teamwork aspects previously discussed with students (e.g., \textit{"[the data] validated some of the points that were made in the debrief"} - T5; \textit{"[the data] highlighted the team communication that we had discussed, and only discussions that we previously had with the group"}, - T4. The debrief tool was also used to \textbf{reflect on a team's reaction in critical moments}, as mentioned by T2: \textit{"for example, from students' positioning, I could see how they reacted, after the MET [medical emergency team] call, if they took a 2-2 approach or a 3-1 approach"}. Teachers also stated that the visualisations were helpful during the debrief because they showed visual evidence of the team's dynamics and performance (e.g., \textit{"it provided visual evidence of team dynamics"} - T7; \textit{"it helped to give visual confirmation of excellent collaboration the team showed"}- T1). All interviewed senior teachers (T1-T4) highlighted how the tool was used to provide what they referred to as "objective feedback", explaining that teachers often focus on one student when observing the simulation, which may later introduce bias towards that student during the debrief (we discuss perceived "objectivity" in the next section on data trustworthiness). For two teachers \textit{"the data [visualisation] was way less subjective"} (T3 and T4). 
The same teachers (T3 and T4) also mentioned how the tool could benefit students' learning by providing alternative sources for reflection (e.g., \textit{"another way of telling students what they did"} - T3; \textit{"another way of seeing their performance"} - T4) and by highlighting positive aspects of students' performance, especially for cases when students assess or judge their own performance too harshly (e.g., \textit{"I highlighted to students they did a good job although they thought had done poorly"} - T4). 

In sum, from the first iteration (that involved data collection only) we learnt that when we clearly communicate the educational purpose of the study to students, they may see the value of participating in a complex MMLA intervention even if they do not receive direct benefits from it (e.g., to help improve current teaching practices or improve the technology that will be used by other students in the future). Moreover, from the second iteration, we learnt that teachers perceived the potential benefits of the richness of multimodal data, rendered into data visualisations, to support students' reflection but they need to ultimately develop the strategies to optimise the effective use of data for educational purposes.  

\subsubsection{\textbf{Data incompleteness and trustworthiness}}
This topic focuses on the potential impact of incomplete multimodal data on the perceived trustworthiness of the MMLA system gathered from students and teachers. When students from the first iteration were asked about trust of their own data presented to them during the interview, 15 out of 20 students indicated that they would \textbf{trust the multimodal information}. Their responses can be organised into two groups. Students in the first group (9 students out of 15) would trust the data because they were able to identify themselves or their perceived team outcomes from the visualisations. For example, S2 explained the following: \emph{``I trust the data because it presents what we did in the simulation, by looking at the data I can see who I was in that scenario''}. Likewise, student S3O indicated the following: \emph{"when I looked at the [visualisations], I could tell that this is what I did, at this time [pointing the visuals], and the actions we performed were very familiar"}. The second group of students (6 students out of 15) responses indicated that they trusted in the multimodal data based on the belief that sensors provide, what they referred to as: ''objective and accurate data''. For instance, one of the students described this as follows: \emph{``I wore the sensor so I know that the information came from pretty reliable sources''} (S3). However, 4 of the 5 students, who did not trust the data, 
were \textbf{skeptical about the trustworthiness of the multimodal information} because at the beginning they found it difficult to read. For instance,  S10 indicated that he \emph{``do not trust the information much because [the visualisations] are not as simple and straightforward''} (S10). In the same way, S22 suggested the following: \emph{``I think I would trust the data for the most part but I think [other people] would require knowledge about the context as well''}. However, when students got familiarised with the visualisations two of them changed their opinions (N=2). For example, S20 suggested that the visualisation was \emph{``a bit hard to interpret. But, once you sort of look at it, and when you read the data as well, when you combine it all together it is trusted and the data presented definitely makes sense''}. In the end, only 3 students would not trust the multimodal data at all as they considered it did not represent what they did during the simulation (e.g. \emph{``I don't think it was accurate enough considering how my team performed''} - S7).

Similarly to the students in the first iteration, most students who looked at their data in the MMLA dashboard in the second iteration (43 out of the 47 survey respondents) indicated that they would generally trust the mulimodal visual representations. However, some students were also aware that sometimes the visualisations in the MMLA dashboard were generated with incomplete data. For example, four students (N=47) recognised that the \textbf{incompleteness of data affected their sense of trust}. One student explained this, as follows: \textit{"This visualisation allows the students to reflect and be accountable in communication when working in collaborative practice. However, I can only trust to some extent as I know that a line is missing between the student wearing blue colour and the doctor for communication"} (S51). Another student expressed a similar idea: \textit{"I cannot judge my communication from these data as the majority of my communication was between myself and the other primary nurse, and her data was not collected"} (S29). As noted by S46, having incomplete data could hinder the validity of results: \textit{"My only comment would be to ensure that the equipment is working and recording as required, as this, unfortunately, affects the validity of results"}.   
  
The four senior teachers also reflected on the trustworthiness of the information presented in the debrief tool. T2 and T3 pointed out that trust is built over time. T2 expressed this idea as follows: \textit{``For the first week at least, I really didn’t understand [the visualisations] to a degree. Once I knew what [the visualisation] was about, I did trust in it in the later weeks for sure"}. T1 and T3 indicated that trust is not coming from the data but from their confidence about making a good interpretation of the information presented in the tool: \textit{``It’s about how I learned to use the system that changed my trust. It is not that I’m not trusting the data. I am not trusting what I have to say about it; I don’t know what it means"} (T1). Moreover, T3 and T4 mentioned that they \textbf{over-trusted} the information. Even though their understanding did not seem aligned with the data, they wanted to make an explanation (i.e., \textit{``Sometimes I was trying to force [an explanation] to the visualisation but I shouldn’t be because it didn’t make sense to me. I was seeing things that I wasn’t expecting or that I couldn’t explain"} -- T4). 

The teachers indicated their \textbf{strong preference for using the debrief tool when the data was complete } (i.e., data from \textit{all four} team members). If the information was about less than four team members, as it was the case for teams in which not all members consented to participate in the study, using the tool can still be relevant. However, teachers had concerns about data incompleteness and a deeper explanation in the debrief was needed to complete the whole picture of what was expected from an entire team according to the expected learning outcomes. T1 explained this concern as follows: \textit{``If [teachers] don’t go through the [incomplete information], [students] don’t understand what they’re looking at, and the wider context. So I walked through how the visualisation would have looked like if we had a full set of students"}. T2 also stressed that visualisations generated from the incomplete data of a team might cause confusion to students: \textit{``giving [incomplete information] to the students without explaining it probably would have just made them confused. I think with three or four [team members] the debrief tool was generally well used"}. But also using incomplete data could be harmful in general as teachers could also make wrong assumptions as T2 explained: \textit{``Sometimes the data would confusing to me. A student reminded me that there were only three [team members] then I was 'yes, that does explain it exactly' "}. 

In sum, we learnt that there may be several risks associated with both students' and teachers' overconfidence in trusting the data merely because it is being automatically captured via sensors. Moreover, teachers and students may also believe that sensor data is objective and free from potential bias. Yet, learning data is intrinsically incomplete \cite{Kitto18Imperfection}, especially when captured via sensors, and data representations are not necessarily objective as they are unavoidably imbued with subjective design decisions \cite{Benford}. It is therefore critical to develop technical and social mechanisms to ensure MMLA systems are trustworthy, which requires teachers and students to acquire a suitable level of understanding that multimodal data is incomplete and subject to different kinds of bias. That incompleteness and bias is both \textit{intrinsic} to computational modelling of any social activity (i.e., all data and algorithms provide partial lenses onto human activity), but \textit{extrinsic} local contextual factors may also be in play (e.g., technical malfunctions; students not consenting to being tracked).

\subsubsection{\textbf{Emerging issues related to visualising multimodal data}}
This topic focuses on issues experienced by students and teachers while interacting with the MMLA dashboard. A recurrent challenge highlighted by all the senior teachers and some of the other teachers (T5, T7 and T9) was that \textbf{they did not feel in the position to make a rapid interpretation of the visualisations} and lead a reflection immediately after. Teachers expressed that this was often due to the challenge of aligning what they observed in the simulation and the information presented in the dashboard. As described by T2: \textit{``sometimes I tried to make sense and align the observed behaviours with the data, but then, during the debrief, I was thinking, Oh no, I don’t think this might be correct"}. Teachers also indicated that they needed more time to familiarise themselves with the MMLA dashboard in order to devise ways to use it in the moderation of the class reflection (e.g., \textit{``it was challenging because I have not done it before”} -- T7, and \textit{``it was challenging due to unfamiliarity with the technology”} -- T9). However, teachers recognised that using the MMLA dashboard involves a learning curve process: \textit{``once we got more familiarised with the tool, it got easier”} (T3). One potential solution is to factor in a preparation or formal training period with the whole teaching team for them to create strategies that they can follow to interpret the data visualisations generated immediately after each team session, as suggested by T4: \textit{``having some standard [moderation] with the teachers would be really useful"}. Another solution suggested by T3 was to enrich the MMLA dashboard with teachers' observations and learning expectations to create a stronger sense of trustworthiness: \textit{``I think we can articulate the data better […] having some kind of moderation where we [the teachers] manually check if students achieved a learning outcome and transfer this into the learning tool"}. 

Some teachers also mentioned that their lack of understanding was due to not knowing how the data was captured and translated into the visualisations (e.g., \emph{``how are you coming up with the yellow bar? What is being used to create that?”} - T4). The four senior teachers suggested that helping teachers to develop their \textbf{data literacy in relation to the particular MMLA dashboard} would contribute to interpreting the visualisations better and also to understanding their limitations, especially when using it for the first time. T1 explained this as follows: \textit{``knowing where the data is coming from and how the information is being calculated would help us to understand the specific parts of it”}. 

Similarly, some students in the second iteration also pointed at the challenge of connecting the complex multimodal data with specific learning constructs or performance metrics. Five out of the 32 students (N=47) who indicated that the information in the MMLA dashboard was relatively easy to read, also explained that it was hard for them to make a clear connection of its meaning with their learning experience  (e.g., \textit{"I understand what each part of the data is showing. But I think I need more explanation to understand the meaning behind each one and the reasoning on why this occurs"} - S54). One student highlighted \textbf{the need for contextual information for them to be in the position of interpreting the multimodal data}: \textit{“I think more of a prompt before the visualisation is shown, such as knowing how many patients we have and the critical information of the patient, then I could have more understanding and confidence of my performance in the simulation”} (S47). Another student expressed her desire for\textbf{ detailed explanations about the team’s performance}: \textit{“I didn't really understand what I was looking at, at first, and there is still some confusion in the last section about task transition. I don't know if being higher than the average is considered a good or bad thing”} (S64). Another student recognised the need for clear instructions to adjust their practice: \textit{“it would be good to see what you would expect to see from a highly effective team”} (S58).

In sum, teachers may find it hard to interpret MMLA visual interfaces given the complexity of the various sources of the intertwined data underpinning them. For effective in-the-wild deployments, teachers need to be supported to develop relevant data literacy skills to understand the basic inner-workings of the analytics and for them to develop pedagogical strategies around the use of the MMLA systems. Students also emphasised the need for design elements in the MMLA visual interface to \textit{explain} the meaning of the data and for them to understand what were the performance expectations.

\subsection{Design and human-centredness}
This theme included the following two topics: i) human-centred design, teaching and learning; and ii) human-centred design and research innovation.

\subsubsection{\textbf{Human-centred design, teaching and learning}}
This topic focuses on teachers' perceived benefits of being part of a human-centred process in the development of the MMLA system in terms of their teaching practices and students' learning.
The senior teachers felt generally \textbf{satisfied with a design process in which their voices could be reflected in the resulting end-user interface}. 
For example, T1 appreciated the partnership with the researchers throughout the design process, as follows: \textit{``We’re all involved in this because we can see that there are benefits to the students and to the teachers for doing this, that it alleviates how we do our debrief and can potentially change the way we do our sims”}. T3 also recognised the value of bringing different expertise to innovate in terms of technology innovation and pedagogical practice: \textit{``I think by combining our teams we can make something that’s new, and we’ll help students learn”}. T1 and T2 expressed that they felt involved in the design process and acknowledged that their lived experiences and suggestions were considered in the final prototype of the MMLA dashboard: \textit{``[the research team] was very flexible in working with us. They have taken a lot of our suggestions on board”} (T2); and \textit{``We felt we had contributed to it. So I thought the two teams were able to add value, and I found that really satisfying”} (T4). Teachers also felt that both researchers and teachers were working towards the common goal of supporting students' learning. T1 and T2 elaborated on this idea, as follows: \textit{``It’s good that we are going towards the same goal, but with different perspectives"} (T2); \textit{``I think it’s important that we’ve been able to have those priorities to see that there’s a common goal. We just want to work at it from different points of view"} (T1). In short, teachers appreciated partnering with researchers in the MMLA design process. This can lead to creating MMLA visual interfaces aligned with existing teaching practices and learning goals. 

\subsubsection{\textbf{Human-centred design and research innovation}}
This topic focuses on the views of both teachers and researchers on partnering to foster MMLA research innovation.
The lead senior teacher (T1) expressed that \textbf{at the beginning of the design process the teaching team was a bit uncertain about the outcome} of the MMLA deployment given its novelty, but the experience after the two MMLA iterations made these feelings fade: \textit{``I really had no idea exactly what we were doing, and I had to see it and be part of it before I could really understand it”}. Linked to this idea, T4 explained that, because the research is considerably innovative, teachers' involvement in the design process enabled them to consider students’ feelings and reactions and be careful about how students were invited to be part of the research study: \textit{``[the deployment] was very innovative. Because of this, I think I was a bit more worried about the [students], thinking that we have to do it correctly. We were given the opportunity to be diligent about it.”}

Also, \textbf{teachers value research collaboration if the common goal benefits students' learning}. T2 expressed this positive relationship by highlighting the openness, easiness and willingness of the research team to work with the nursing teaching team: \textit{``It’s the fact that there are some researchers who are just as passionate about this as the educators, it’s actually easy for us to keep wanting to do this because they want to work with us as much as we want to work with them. And although they aren’t nurses, actually we can work together and share points of view, so that we can try and understand what the research side is. They are willing to work, and be quite flexible with what’s going on, so that we can make it work for nursing"}. T3 also stressed the strengths of combining both, the LA research and the nursing teaching teams towards building innovative learning tools: \textit{``I think by combining our teams, we can make something that’s new, and you know we’ll help students learn"}.

Researchers' perspectives resonated with teachers' perspectives. The research team indicated that it was critical to consider \textbf{teachers' voice since they have the lived experience in deploying pedagogical interventions}. For example, T4 described how teachers would indicate whether certain logistic decisions were \textit{``unfeasible"} or not. R4 explained this as follows: \textit{``When collaborating or interacting with teachers, I think the most important thing is to put teachers' and students' educational needs before our research desires”}. R3 also explained that a benefit of collaborating with teachers in the design of the MMLA deployment was that \textit{``they could be invited to an interview, to get what they valued or missed in the visualisation which can show what can be improved or developed in the future to help them better"}. R4 also explained how important was to collaborate with teachers to \textbf{understand the meaning of the data }in light of the characteristics of the learning design: \textit{``This collaboration has led to changes in my algorithm for detecting task prioritisation. There are some minor changes in the learning design that we were unaware of but significantly impacted the analytics"}. Finally, R1 also mentioned the importance of \textbf{partnering with students} to validate the MMLA dashboard: \textit{``Doing a cognitive walkthrough while teachers and students think aloud has benefited my research in validating how useful or misleading the [multimodal] visualisation may be"}.

In sum, we learnt about the importance of involving teachers and students in the design process to not only validate the highly innovative MMLA end-user interfaces but also to expand understanding of the learning design, values that must be endorsed, and the lived experiences that can affect the logistics of the deployment. 

\subsection{Social Factors}
This theme included the following two topics: i) consenting and participation strategies; and ii) data privacy and sharing.

\subsubsection{Consenting and participation strategies}
This topic focuses on lessons learnt from the consenting strategies applied in both iterations of the study. Teachers recognised that, proportionally, more students consented to participate in the first iteration (2021) than in the second (2022). As mentioned by T4: \textit{``it didn’t seem to work as well [the second] time. We were trying very much to keep on time, and I think [the consenting process] actually became a little bit more complicated”}. While some teachers tried to explore potential explanations for this difference, citing potential seasonal differences (e.g., \textit{``students may just be so much more burnt out than they were last year. They’re disengaged on everything, not just the simulation”} -- T1); teachers tried to optimise the consenting process. In preparation for the second iteration, consenting information was sent to students beforehand as pre-class material through an online consent form and an explanatory video. Teachers recognised this strategy may have caused confusion since students often do not access pre-class materials, as explained by T4: \textit{``we also know that they don’t access the pre-class stuff, so that will limit their exposure to it. But I did feel it was just a little bit complicated for them to take on, and then agreed to consent to”}. T3 also explained that \textbf{the MMLA deployment may be too novel in the eyes of the students} and it involves several layers in relation to data (i.e., capture, analysis and visualisation) that students cannot fully comprehend:\textit{``I think the video was really helpful. We played it in the class, and it was very sound, very personable and short. Perhaps the way it communicated made it sound a bit complicated”}.

Due to the challenges in getting participants for the second data collection, the senior teaching team shifted to the same strategy as in the first iteration (2021). This consisted in inviting a researcher to give a short 1-2 minute speech just before the class to communicate the research and ask them to participate. Students could ask any clarification questions. As mentioned by T1: \textit{``I think it worked really well when one of the researchers gave the speech from the research point of view. And then one of the academics back that up”}. This space also served to explain to students that the data was de-identified and that participating in the research would not affect their grades. As mentioned by T2, this strategy was crucial to highlight to students that the data was being de-identified: \textit{``We noticed that they were worried like if their faces or names will be shown in the screen”}.  However, packing all ideas in two minutes is challenging. It may cause some pressure on the research team. T4 indicated that: \textit{``you’re trying to [inform students] in a very short space coming in and trying to convey this information. So I think [the researcher] was under a lot of pressure to do it”} (T4). 

Moreover, all the senior teaching team agreed that the consenting strategy should be improved. After teachers experienced two in-the-wild MMLA deployments, they reflected on the value of the research, the usefulness of the MMLA debrief tool and its potential impact on students’ learning. They thus indicated they may want to move the deployment forward to happen as a part of the regular tasks in the classroom. They proposed a \textbf{'business-as-usual' use} of the MMLA dashboard if the maturity of the system allows it as in video-based debriefs in healthcare simulation \citep{megel2013high}, so all students would have the same learning opportunities. This way, the consenting would be just about optionally \textit{recording} the data for research purposes, otherwise, deleting it immediately after the class. This was explained by T1 as follows: \textit{``I still think our best bet is that the [MMLA system] is inbuilt in all simulations. Perhaps this is what we will do next year. Maybe we have only one or two sessions where the technology is not used at all. This is definitely the way to go}".

R3 and R5 agreed that the lower participation in the study was due to the intrinsic complexity of the multimodality aspects of the data collection. R3 expressed that \textit{``some students didn't seem to actually understand what would happen in our data collection and didn't want to participate. Students might see the digital version of the consenting form as more work and then ignore that"}.  The lack of understanding could come from technical words embedded during the explanation of the study. R5 indicated that \textit{``[students] didn't actually understand what’s 'multimodal data' and didn't ask about it either”.} As reflected by R5, there should be a balance between details (e.g., over-explaining details of the data and technology that students without formal analysis training and AI literacy would not easily understand ) and simplicity (e.g. omitting key information about the potential implications related to the data collection that may be relevant to make an informed decision) to minimise students misunderstandings: \textit{``over-explaining technologies, explaining from what kind of data and showing what kind of visualisations they’ll see in debriefing, resulted in a lower number of students participating in the study. However, explaining it too simply and straightforward resulted in several students withdrawing from the study”}. For future studies, researchers suggested \textbf{simplifying the explanations and words} in the consent form and during the description of the research. R5 explained this as follows: \textit{``I reckon we need to improve some wording in the consent form by simplifying some words, for example, instead of using the word 'multimodal' we should use words they can understand such as 'audio, position, and health data')}.



When reflecting on consenting strategies from both iterations, R5 indicated that consenting was better in the first iteration because it was \textbf{focused only on data collection}. Therefore, teachers were not much involved as in the second iteration: \textit{``The consenting [from the previous year] was smoother than this year because we only focused on data collection. Teachers didn’t provide much help besides helping students wear sensors (i.e., pozyx, mic, and empatica)”.} R4 also reflected that \textit{``although the consenting strategy in the first year was non-environmentally friendly, as printed consent forms were handed to each student, it was very successful in getting several students volunteering to the study”}. It seems that an in-person strategy to explain to students about the MMLA deployment is better than an online strategy because clarifications can be made in person, and there is a direct engagement with students: "an online strategy can be confusing because students do not only consent to the study but also to the online confidentiality agreement they need to sign as a part of their course" (R3).

In sum, it is challenging to explain to students what a complex MMLA study entails as it involves various types of heterogeneous data sources each pointing at multiple analysis and visualisation approaches. Yet, providing too many technical details about the sensors and the analytics in advance may not necessarily contribute to clarity. Explaining the complexity of the MMLA deployment \textit{in person} can enable students to ask clarification questions and then provide informed consent. 


\subsubsection{Data privacy and sharing}
This topic focuses on students' perspectives on multimodal data privacy and sharing. 
All interviewed students (N=20) agreed to \textbf{share their data for educational purposes in ways they could not be identified }by others outside their class. Yet, they had various different perspectives about who should benefit from looking at and using the visualisations. For instance, 10 students indicated that it would be only beneficial for those students who own the multimodal data to use it. As one of the students argued: \emph{``you just take more knowledge from your own experience than with someone else's experience''} (S5). This was supported by another student who considered that showing their own data \emph{``will make more sense to the team, who was there performing the simulation''} (S18). The rest of the students (N=10) reflected on the opportunities of sharing the data with others and its benefits for learning. For instance, S20 envisaged that \emph{``if [he] could see their teams' visualisations, compared to another team that did the exact same scenario, it would be quite interesting to see maybe why did this team recognise aspects earlier than them''}. Likewise, other students (N=5) indicated that teachers can benefit in the way they \emph{``can work on improving critical aspects of students performance and the scenario''} (S10), \emph{``can learn a lot about student thinking and performance''} (S18), and \emph{``can help them understand how well a student did''} (S3).

In sum, students did not show signs of being concerned about sharing their data with others, which may be explained by a low level of understanding of the multimodal data itself as mentioned in the previous subsection. Yet, half of the interviewed students thought that the data would be meaningful and relevant only to the students who were in the same class where the data was collected. 

\subsection{Sustainability} 
This theme included the following two topics: i) technological sustainability; and ii) MMLA appropriation in the classroom.

\subsubsection{Technological sustainability}
The first topic relates to the sustainability of the technological infrastructure. 
R1 and R5 suggested a flexible `detachable' architecture capable of \textbf{running through microservices as a centralised process}. The idea of this architecture is to minimise data loss in case one of the data sources is not correctly running (e.g. “\textit{the architecture should provide an easy way to de-attached pieces of the application presenting issues. That way, it would be possible to reduce any misleading data visualisation [caused by issues during data collection]”}, R1). Microservices should also allow running the MMLA solution with minimum requirements: \textit{“At a minimum, these systems should be able to run with minimum hardware and software installations. For example, we have one document that provides a list of recommended hardware that we use (i.e., audio interfaces, pozyx, etc..) and a list of software that is needed to be installed (such as ASIO4All for audio and Pozyx interface)” }(R5). 

Special care should also be given to automatically \textbf{communicate to teachers about any issues that may have been detected during the data collection} (e.g., sensors that may have been accidentally disconnected) so they can decide if they will use the tool in the debrief or not: \textit{“teachers should be informed that issues can happen and it can help them to decide whether they want to use the technology or not during their classes”} (R1). R4 also explained that technological sustainability depends on the readiness of sensing technologies, switching towards user-friendly versions of the hardware: \textit{“The sustainability of MMLA research studies largely depends on the readiness of the sensing technologies. For example, the positioning tracking system and analytics will become sustainable for teachers to use on their own when fully enterprise solutions are available, as the level of automation from data collection to analytics will be higher”}.
In short, a potential strategy to maximise long-term technical sustainability is a light-weight microservices-based architecture that can enable attaching and detaching heterogeneous sensors as required. Such technical properties of the MMLA infrastructure are tied to \textbf{building and sustaining stakeholders' trust} in it. 

\subsubsection{MMLA appropriation in the classroom}
The second topic concerns the requirements to sustainably integrate the MMLA system into regular teachers' practices. 
R1 and R2 expressed that for teachers to appropriate the MMLA system, this should move towards a toolkit (of hardware and software) that can be easily deployed and \textbf{used by non-experts users}, such as the teaching team or nursing students. R2 suggested that \textit{"all the software should be reconstructed to generate a toolkit which can be easily deployed and used”}. R1 further expanded on strategies required to embed the multimodal sensors into the existing classroom ecology, as follows: \textit{“a classroom can be equipped with charging stations where teachers and students can collect/leave the devices they are wearing, and teachers have a computer in the classroom where they can easily start/stop the data collection and generate data visualisations}". 

The senior teachers' perspectives were also aligned with the idea of running the MMLA system without requiring too much technical support: \textit{“in terms of using the system, running the tool by yourself without any help of the team of researchers. Right? I think that would be the ideal end goal that we can just run it.”} (T2); and \textit{“there should be some way of being able for it to collect it [data] automatically”} (T4). However, both T2 and T4 also suggested that\textbf{ minimal technical support is still required}: \textit{“it will always require a sort of a technical person”} (T4) and \textit{“if we get to the point where we only need one researcher there on the day to troubleshoot, just in case things go wrong to get that [the system] up”} (T2). R1 mentioned that \textit{“having a dedicated space (e.g., a specific classroom) to run this multimodal data collection on a regular basis would help other universities if they want to implement these solutions)”}. T4 supported this idea: \textit{“concerning the technology, I think that the infrastructure in the room could be improved, the sensors need to be embedded in the room”}. R3 mentioned that it would be ideal to use the equipment that is already part of the simulation room, such as ceiling microphones and 360 video cameras: \textit{“If we can use the devices they already have, like microphones, the price of the whole system would be cheaper”}.

Finally, researchers and teachers suggested \textbf{a training period for teachers}. For example, T4 explained the following: \textit{“teachers could definitely be better educated about it and be more autonomous in that regard. It is about training the teaching staff to be able to attach the wearables and all the rest or running this the MMLA system”}. 
In sum, a potential strategy to maximise adoption and technology appropriation by non-technical end-users includes i) embedding sensing capabilities into the classroom; ii) providing teachers with a high degree of user control; iii) providing basic training for teachers to not only use the system but also to learn how to interpret and act upon the multimodal data representations effectively; and iv) providing readily available technical support in case something goes wrong.

\section{Discussion}
In this section, we summarise the lessons learnt from our MMLA in-the-wild deployment; then discuss the implications of these findings for practice, identify various limitations of our in-the-wild study, and suggest some potential directions for future research and development.

\subsection{Summary of lessons learnt}

This paper provides a summary of some of the key logistical, privacy and ethical challenges that emerged from our complex MMLA, in-the-wild study. These can be listed as follows: \hfill \break

\emph{Space and place}
\begin{itemize}
\item \textbf{Intrusiveness} -- While students did not report discomfort in wearing sensors, teachers can still get concerned about their potential \textit{distracting factor} and some students can feel \textit{stressed} about being monitored. 

\item \textbf{MMLA Technology readiness} -- The lack of MMLA technology readiness can severely impact the lesson plan. Teachers need to play an active role to create \textit{strategies to moderate} the sensing/analytics technologies, and minimise potential disruptions and setup time.  

\item \textbf{Unexpected issues during the MMLA deployment} -- While several technical issues that can emerge during the MMLA deployment are beyond the control of the research team, reducing the number of devices used can minimise potential technical failures. Some high-end sensors may need to be replaced with less expensive sensors, that may capture coarser data, if the change increases \textit{reliability}. 

\item \textbf{Multimodal data quality, portability of sensors and affordability} -- At least currently, a trade-off may exist between capturing \textit{high quality} data and the portability and affordability of the sensing technology.
\end{itemize}

\emph{Technology: data and analytics}
\begin{itemize}
\item \textbf{Purpose of capturing multimodal data} -- If communicated clearly, students are willing to participate in a complex MMLA study and contribute their data for the purpose of helping their teachers or future students. Teachers can and need to develop strategies to optimise the use of multimodal data to support students.  

\item \textbf{Multimodal data incompleteness and trustworthiness} -- Although multimodal data is required to build analytical representations of an embodied learning experience, multimodal sensor data are intrinsically incomplete and subject to bias. Thus, mechanisms to ensure MMLA systems are \textit{trustworthy} and designing for data incompleteness are required. 

\item \textbf{Emerging issues related to visualising multimodal data} -- Teachers need to be supported to develop relevant \textit{data literacy skills} to understand the basic inner-workings of specific MMLA systems and for them to develop pedagogical \textit{strategies around the effective use} of the intrinsically complex MMLA visual interfaces. Students may also require visualisation guidance or explanatory features for them to the meaning of the data in educational terms.
\end{itemize}

\emph{Design: human-centredness}
\begin{itemize}
\item \textbf{Human-centred MMLA and students' learning} -- Teachers' appreciation of partnering with researchers in the design process can lead to creating MMLA systems aligned with teaching practices and learning goals. 

\item \textbf{Human-centred MMLA and research innovation} -- Involving teachers and students in the design process contributes to the validation of the MMLA interfaces according to the learning design and to the improvement of the logistics of the MMLA research study. 
\end{itemize}

\emph{Social factors}
\begin{itemize}
\item \textbf{Consenting and participation strategies} -- It is challenging to explain to students what a complex MMLA study entails. Providing too many technical details about the sensors and the analytics in advance does not necessarily contribute to clarity. Explaining the complexity of the MMLA deployment \textit{in person} can enable students to ask clarification questions and then provide informed consent.  

\item \textbf{Data privacy and sharing} -- Students were willing to share their multimodal data with others if their privacy is preserved and the purpose is limited to supporting learning. While most students see their multimodal data as only beneficial to themselves, some students can see the potential benefit to make their data available to other students to learn from their experiences or for teachers to improve the design of the learning tasks. 
\end{itemize}

\emph{Sustainability}
\begin{itemize}
\item \textbf{Technological sustainability} -- A potential strategy to maximise long-term technical sustainability is a lightweight \textit{microservices-based architecture} that can enable attaching and detaching heterogeneous sensors as required.

\item \textbf{MMLA appropriation in the classroom} -- A potential strategy to maximise adoption and technology appropriation includes embedding sensing capabilities into the classroom, providing a high degree of user control, providing training to teachers on system usage and data interpretation, and keeping the need for support from a technical actor to a minimum extent.
\end{itemize}

\subsection{Implications for practice}
The lessons learnt from our in-the-wild MMLA study have several implications. We summarise these into the following three recommendations to provide guidance for researchers, developers and designers to make informed decisions about the effective deployment of MMLA in-the-wild. 

\textbf{\textit{Forging design partnerships with teachers and students}.} The more sensors are used to capture activity in complex educational scenarios that involve non-computer mediated interactions, or ill-defined, open tasks such as in teamwork, the more complex the meaning-making process becomes to move from data to insights  \citep{echeverria19towards}. Thus, as rich data infrastructures become more commonplace in educational contexts \citep{guzman2021learning}, it is also becoming critical to forge strong partnership relationships among teachers, students, educational decision-makers, researchers and developers. This has the potential to ensure that algorithmic outputs and data representations are meaningful and aligned to local learning objectives and pedagogical values \citep{Ahn2019}. Indeed, some educational researchers have started to utilise the body of knowledge and practice from design communities, such as participatory design and co-design, in data-intensive educational contexts \cite{BuckinghamShum2019}. However, following human-centred design approaches is yet to be seen in MMLA according to the most recent review \citep{yan2022scalability}. 

In our study, several practical challenges in the MMLA deployment demanded expertise from a wide range of areas (such as learning analytics, interaction design, and information visualisation), plus knowledge from stakeholders contributing insights and evidence from their lived experiences. By giving an active voice to students and involving teachers in the design process we were able to identify the key practical challenges that can easily undermine adoption if they are not addressed in a timely manner. Teacher/student involvement was also critical to give meaning to the complex multimodal data streams both for research purposes, and to design the MMLA dashboard aimed at end-users. An indicator of the success of the teachers' partnering experience, is that once they reflected on the value of the MMLA deployment, they wanted to move the deployment to happen as a part of their regular classes, potentially making the transition from research to practice an immediate possibility. 

Yet, much work is still required to develop specific guidelines to create human-centred MMLA systems. For example, the rapidly growing human-centred AI \citep{shneiderman2021human} movement within and beyond HCI has much to offer to the design and development of MMLA systems to ensure that novel AI tools are effectively in service of students and teachers. Moreover, researchers and developers may want to address the complexity of visual interfaces of multimodal data by grounding their designs in key Information Visualisation principles aimed at scaffolding the interpretation of large amounts of data by non-technical users (e.g., by applying data visualisation guidance \citep{ceneda2016characterizing} or data storytelling \citep{martinez20} principles).

\textbf{\textit{Designing MMLA considering data imperfection and teacher control}.} 
In Jeffrey Heer's view \citep{heer2019agency}, \textit{"AI methods can be applied to helpfully reshape, rather than replace, human labor"}. In our study, the ultimate aim is not to replace the teacher but augment their repertoire of tools they can use to support students' reflective thinking through data interfaces. Yet, the data captured from the physical world through sensing devices are often incomplete, noisy, and unreliable \citep{bamgboye2018towards}. Moreover, beyond the use of multimodal data in education, it has been  reported that there is commonly a disconnection between logged data and higher-order educational constructs \citep{echeverria19towards, mangaroska2018learning}. This means that the design of effective MMLA interfaces needs to deal with data incompleteness and partial models of the actual learning activity. Creating MMLA systems that perform fully automated actions based on these incomplete data can thus be risky, and cannot be recommended at this level of MMLA maturity.

A primary finding from our MMLA in-the-wild study is that teachers see that a key requirement to maximise the sustainability of the complex computational system is to provide a high degree of user control. The debate around the balance between human agency and AI automation is not new in HCI \citep[e.g.][]{shneiderman1997direct}, yet, it is nascent in the context of MMLA. Nonetheless, \citet{Ogan19} suggested that once sensing technologies mature to the extent that they enable capturing a variety of behaviours in the classroom, we should let teachers empower themselves to use data for making informed decisions and improving their own classroom practices. 

Moreover, we learnt that if the MMLA interface does not provide any visual cue about potential data incompleteness, both teachers and students can attempt to make potentially misleading inferences from the data. More problematically, decisions can be made and actions can be taken without sufficient recognition that logged student data is, by definition, imperfect \citep{Kitto18Imperfection}. In the long term, this can damage their trust in the system. 

Future work can consider at least two potential ways to address these challenges. First, as suggested by some of the teachers in our study, it may be possible to identify gaps in teachers' knowledge around the use of data in their practice such as whether they are aware of how the multimodal data are collected, what educational constructs are being modelled, the limitations of algorithmic outputs, and the kinds of insights that can be derived from them. Professional development programs can be created to increase teachers' AI literacy \citep{long2020ai} and visualisation literacy \citep{pozd2023} for them to understand, to some extent, how they can integrate the MMLA interfaces into their existing practices or how they can adapt their current practices to the new possibilities enabled by the use of such multimodal data. Alternatively or in parallel, the teachers in our study also suggested that the MMLA user interface can be designed to provide visual cues that alert teachers about the reliability of the data so they can make informed data interpretations or decide not to use the MMLA system for a session with uncertain data. To address this, researchers and developers of this kind of innovations may want to consider elements from the emerging literature on the human aspects of AI explainability \citep{JIANG2022102839,khosravi2022explainable} to design MMLA systems that, for example, reveal their assumptions and biases in ways that make sense to non-specialist users so they can keep in control of the potential pedagogical actions that can be taken \citep{selwyn2019s}.

\textbf{\textit{Ensuring teachers' and students' safety}. }
Enhancing physical learning spaces with rich sensing capabilities unavoidably raises critical questions about the potentially harmful effects of excessive surveillance and potential threats to students' and teachers' privacy rather than supporting learning. Preserving human safety in increasingly autonomous smart environments has been identified as one of the main HCI grand challenges \citep{stephanidis2019seven}. \citet{selwyn2019s} explains that even learning analytics systems intended to only support students' learning run the risk of being utilised for broader purposes: \textit{"the concern here lies with the secondary (re)uses of learning analytics data by institutions and other `third parties'”} (p.3). Multimodal learning data can raise particular concerns since analysing a combination of on-skin and under-skin sensor data can lead to richer user models that could be used for student profiling or for performance measurement of teachers which may have negative consequences for the individuals concerned \citep{selwyn2018doing}. Unfortunately, the ethical implications of using MMLA systems have been seldom mentioned in the literature, as has been flagged in recent scoping works \citep{cukurova2020promise,worsley2021new} and reviews \citep{Alwahaby2022,crescenzi2020multimodal}. 

 Our findings flagged some further concerns. Teachers and students may not easily grasp all the potential ways in which their data can be exploited. Yet, they had sufficient awareness to confirm that their data should only be used by themselves or by other educational stakeholders to support other students. Strict guidelines about data privacy and data ownership should be established for systems that use students' multimodal data since some of these data can be highly sensitive. For example, designers could explore ways in which end-users can indicate to the MMLA system to forget their multimodal data totally or partially after it has been used for educational purposes \citep{Muller2022}. Visualising multimodal data also raised another set of potential concerns. In our second iteration, students' inclinations to participate in a MMLA study changed as they seemed to be more willing to participate in a study that only involved data collection but were not sure about all the implications related to having a user interface showing their data in front of their classmates. In this regard, future MMLA work aimed at closing the learning analytics loop by providing end-user data interfaces would benefit from building upon the long-standing HCI research focused on designing for sharing personal data through group interfaces \citep{greenberg1999pdas}. Moreover, although some preliminary work has attempted to discuss ways to effectively write up consent forms for MMLA studies \citep{beardsley2020enhancing}, further work is needed to understand how students can make informed decisions regarding their participation in MMLA studies or in terms of data sharing based on the types of data used in a particular MMLA innovation. 


\subsection{Limitations}
Our study has various limitations. First, the lessons learnt are not generalisable as MMLA studies cannot be treated as a generic type of analytics. Our study involved the use of video, physiological wristbands, audio, and indoor positioning sensing. Although these cover most types of sensors used in  MMLA studies \citep{yan2022scalability}, students' and teachers' perceptions towards sensing technologies can vary across learning situations and technical setups. For example, in other studies where laboratory-grade EEG headsets have been worn by students, their perceptions towards potential negative effects related to sensor intrusiveness have been more prominent compared to those of the students in our study \citep{mangaroska2021challenges}. 

A second limitation is that the teachers and students in our study were, to some extent, accustomed to technology-equipped learning spaces, such as the simulation rooms. Thus, our MMLA sensors were added to an existing ecology of devices and educational practices that involve the use of technologies of various kinds. Nonetheless, most of the existing technologies are not used for the purpose of monitoring and data-intensive reflection thus the lived experiences of the educational stakeholders were novel in relation to the MMLA innovation. 

A third limitation is that the students who participated in the study and the interviews were those who were more willing to participate and often highly motivated as participation was optional. We could not interview participants who were less inclined to experience the MMLA study which prevented us from gaining a deeper understanding of the factors considered by non-consenting students or potential further concerns about the deployment. 

Besides the comments from students and teachers, we also reported some of the lessons learnt from a researcher's perspective with the aim of sharing the particular experiences and insights we gained from this in-the-wild experience. Readers are encouraged to interpret these as such rather than as generalisable claims. 

Finally, evidence was captured heterogeneously from iterations 1 and 2 of our study (e.g., students were interviewed about intrusiveness in iteration 1 but not in iteration 2). This was a consequence of conducting the study under authentic conditions in which the research aims adapted to the needs and availability of the teachers, the students and the planned educational activities. Yet, we did not want to challenge these to preserve the in-the-wild nature of the study.

\section{Conclusion}
Multimodal learning analytics have the potential to open exciting new avenues for supporting teachers and learners, with multiple caveats. In this paper, we have explored some of the challenges that emerged in a MMLA study conducted in-the-wild. Some of these can be addressed by improving the design and logistics of the study and adapting current educational practices. Yet, other challenges require a human-centred perspective to design for the ethical use of sensing and analytics technologies. This paper should be seen as an initial stepping stone of much more work aimed at generating a deeper understanding of effective practices to deploy sensing and MMLA technologies in authentic learning situations with integrity.

\begin{acks}
This research was funded partially by the Australian Government through the Australian Research Council (project number DP210100060). Roberto Martinez-Maldonado’s research is partly funded by Jacobs Foundation.
\end{acks}

\bibliographystyle{ACM-Reference-Format}
\bibliography{references}

\appendix

\section{Survey and interview guide}
In this section, we provide the survey questions and interview guide for our study. 

\subsection{Student's interview}
\label{ap:students-interview}
A total of 20 students participated in the interview during Iteration 1 of the study. Students were asked the following questions:
\begin{flushleft}
\textbf{Theme 1: Space and place}

     - What data do you recall was collected during the simulation?
     
     - Why do you think the data was collected during the simulation? 
     
     - How did you feel wearing sensors during the simulation?
     
     - Did you have any concerns about wearing additional sensors and the microphone?

\textbf{Theme 2: Data and Analytics}
  
     - Would you trust the information presented in the visualisations?

\textbf{Theme 4: Social Factors}

     - Who do you think should look at these data?
     
     - Would you be willing to share this data for academic purposes? E.g.: for the teacher to guide the debriefing session?
   
\end{flushleft}

\subsection{ Teaching team survey}
\label{ap:teacher-survey}

\begin{flushleft}
The following survey questions were handled to the teaching team at the end of the debrief session. A total of 11 teachers answered the survey in the Iteration 2 of the study.

\textbf{Theme 2: Data and Analytics}
  
     - How did you integrate the data into the debrief?
     
     - Did you think the data assisted student learning?
     
     - Did you find using the data during the debrief challenging? Why?
   
\end{flushleft}

\subsection{Senior teaching team interview}
\label{ap:teacher-interview}
In this interview, a total of 4 teachers participated in Iteration 2 of the study.
Teachers were asked the following questions:

\begin{flushleft}
\textbf{Theme 1: Space and place}
  
 - How intrusive was it to equip students, teaching staff and the simulation space with various sensors?
 
 - What kind of unexpected (technical and/or logistic) issues did you face that may have affected the learning goals of the simulations?
 
 - How do you think those unexpected issues (technical and/or logistic) can be minimised in the future?
 
 - Do you think you (or other teachers) may need some training with the tool beforehand?

\textbf{Theme 2: Data and Analytics}
  
 - How did you use the “tool” (slides) during the debrief? For what purposes?
 
 - To what extent do you think that using the tool may have been helpful for you or the students during the debrief?
 
 - Do you think any of the data presented may have been misleading? If yes, explain how?

 - Did you trust the information presented to you through the tool during the debrief? 
 
 Did you also use the visualisations for cases where the data were incomplete (for example, when not all the students were tracked)?
 
 - How can we improve the system so you can trust more on the data?

\textbf{Theme 3: Human-Centredness}
  
 - To what extent do you value the collaboration with learning analytics researchers to design this technology with them? What motivates you to do that?

\textbf{Theme 4: Social Factors}
  
 - What do you think about the consenting strategy from last year's study (iteration 1) in comparison to the one for this year's study (iteration 2)?
 
 - What do you think can be done differently regarding the consenting strategies for a future study?
 
 - Did you perceive or hear any concerns from students regarding the study? If so, can you explain?
 
 - Besides the students and the teacher leading the debrief, who do you think would benefit from looking at the visualisations shown in the tool used in the debrief?

\textbf{Theme 5: Sustainability}
  
 - How can we run our research studies in the future in a more sustainable way? 
 
 - What steps would be needed to make our current system into a real-world application without the help of a team of researchers behind it?
   
\end{flushleft}

\subsection{Researchers' survey}
\label{ap:researchers-survey}
At the end of Iteration 2 of the study, five researchers from our team were asked to fill in a survey, comprised of the following questions:

\begin{flushleft}
\textbf{Theme 1: Space and Place}
  
 - How complex was it to transport, install and configure the equipment before the data collection location? 
 
 - Did you face any specific challenges/problems?

 - What kind of unexpected technical challenges/problems did you face regarding the sensors/equipment during the study/data collection?

 - What kind of unexpected logistic issues did you face that could affect the study?

 - How do you think those unexpected issues (technical and/or logistic) can be minimised in the future?

\textbf{Theme 3: Human-Centredness}
  
 - What do you value the most when collaborating or interacting with teachers or students to plan, analyse or validate your research ideas?

\textbf{Theme 4: Social Factors}
  
 - What do you think about the consenting strategy from last year's data collection (iteration 1)? 

 - What do you think about the consenting strategy from this year's data collection (iteration 2)?

 - Please, share your ideas on what you think can be done differently regarding the consenting strategies for a future study/data collection.

\textbf{Theme 5: Sustainability}
  
- How can we run our research studies in the future in a more sustainable way? 

- How can we recover from failure and debug issues to provide reliability to our systems? 

- What steps would be needed to make our current system into a real-world application without the help of a team of researchers behind it?

\end{flushleft}

\subsection{Student's survey}
\label{ap:students-survey}
A total of 47 students completed a survey during Iteration 2 of the study. Students were asked the following questions about the visualisations presented in the MMLA dashboard:
\begin{flushleft}
\textbf{Theme 2: Data and Analytics}

\textbf{\textit{Visualisation 1: Team Communication}}

Please review the following visualisation that represents the data we collected during your simulation. Reflect on what it may represent and answer the questions below. 

\begin{figure}[h!]
\includegraphics[width=12cm]{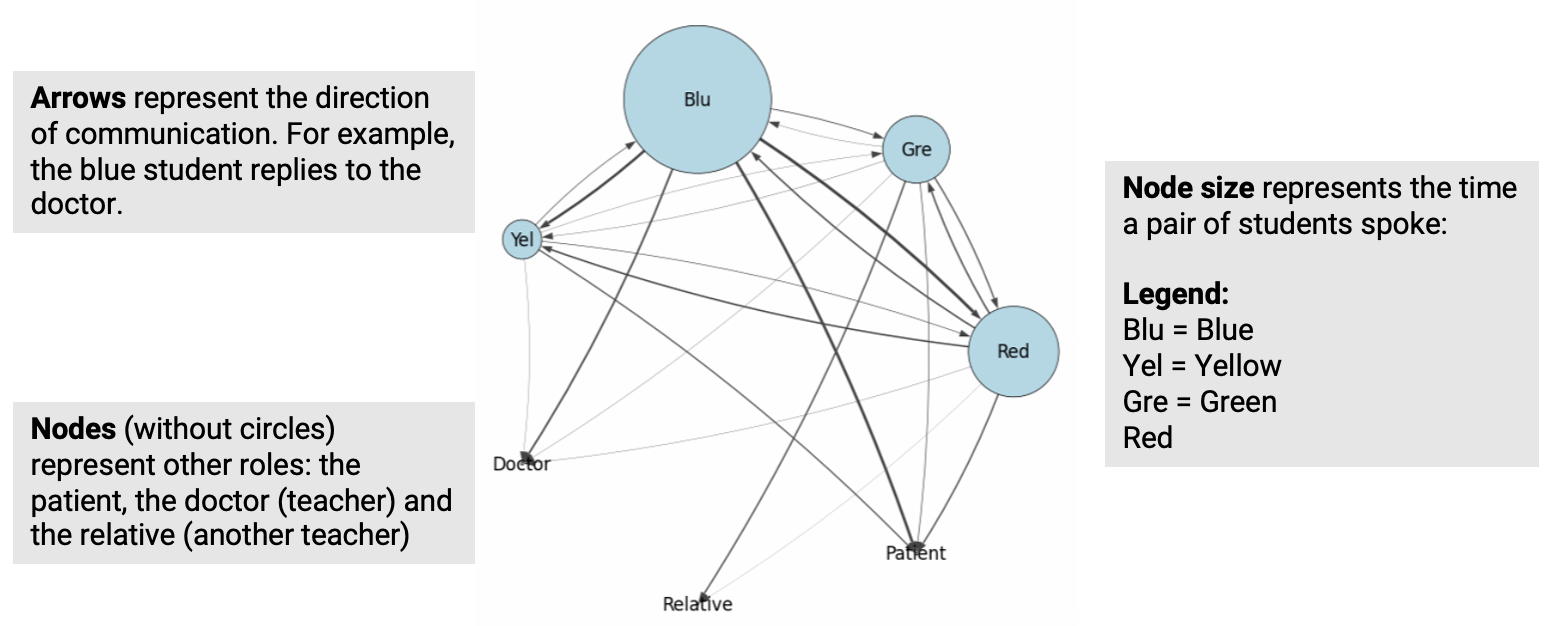}
\end{figure}

\textit{Note: this visualisation IS representing YOUR data}

- To what extent you would trust this visualisation to judge or reflect on your own performance?

        \begin{enumerate}
        \item I’d absolutely trust on it
        
        \item I’d trust it to some extent
        
        \item neutral/borderline
        
        \item I’d not trust it to some extent
        
        \item I’d absolutely not trust on it
        \end{enumerate}
        
- Please, briefly explain your response.

- Do you have any comments on how this visualisation can be improved, or would you add something to the visualisation to make it more straightforward?   
\pagebreak 

\textbf{\textit{Visualisation 2: Team Speaking and Positioning}}

Please review the following visualisation that represents the data we collected during your simulation. Reflect on what it may represent and answer the questions below. 

\begin{figure}[h!]
\includegraphics[width=12cm]{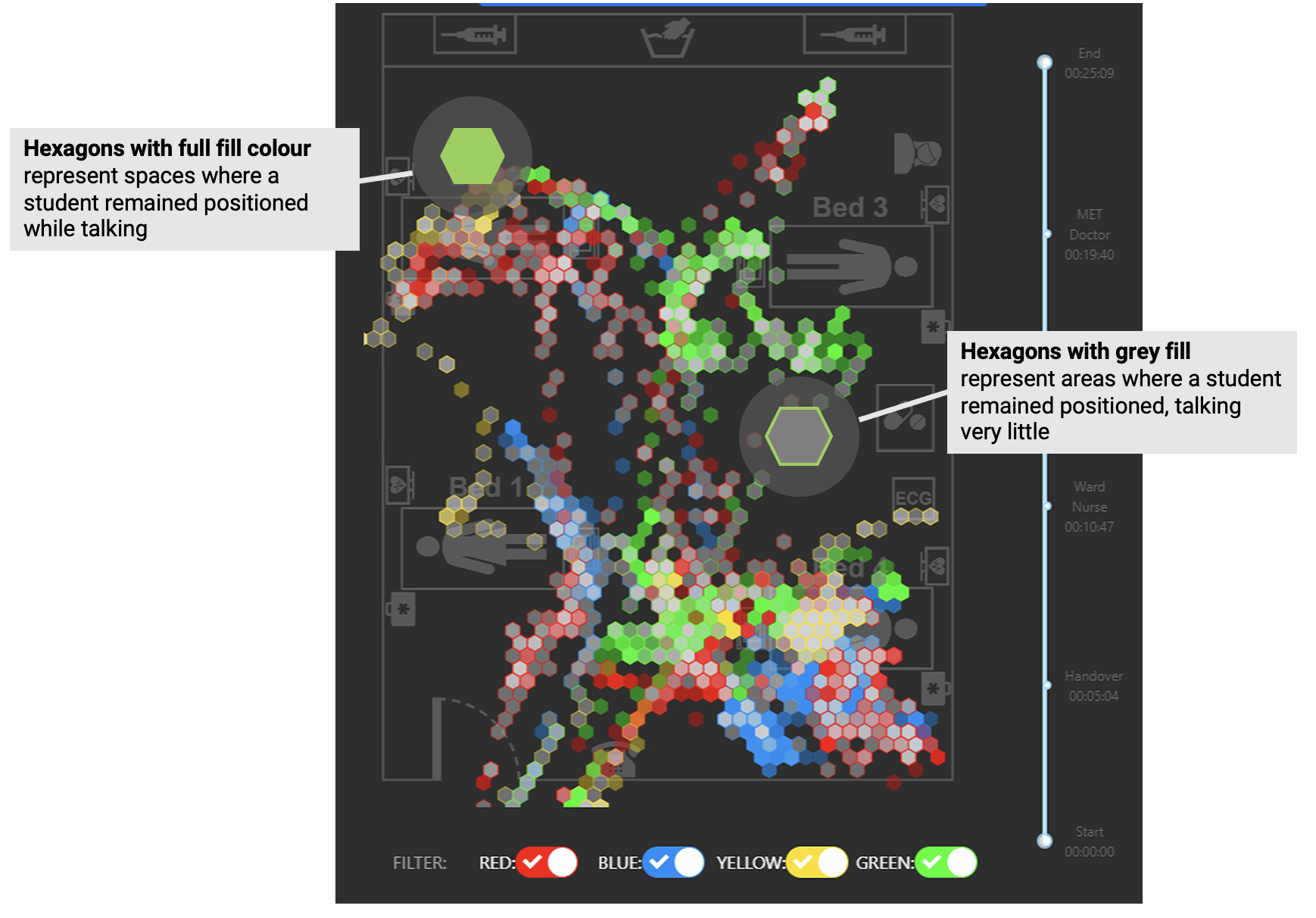}
\end{figure}

\textit{Note: this visualisation IS representing YOUR data}

- To what extent you would trust this visualisation to judge or reflect on your own performance?

        \begin{enumerate}
        \item I’d absolutely trust on it
        
        \item I’d trust it to some extent
        
        \item neutral/borderline
        
        \item I’d not trust it to some extent
        
        \item I’d absolutely not trust on it
        \end{enumerate}
        
- Please, briefly explain your response.

- Do you have any comments on how this visualisation can be improved, or would you add something to the visualisation to make it more straightforward?   
\pagebreak

\textbf{\textit{Visualisation 3: Team Prioritisation}}

Please review the following visualisation that represents the data we collected during your simulation. Reflect on what it may represent and answer the questions below. 

\begin{figure}[h!]
\includegraphics[width=12cm]{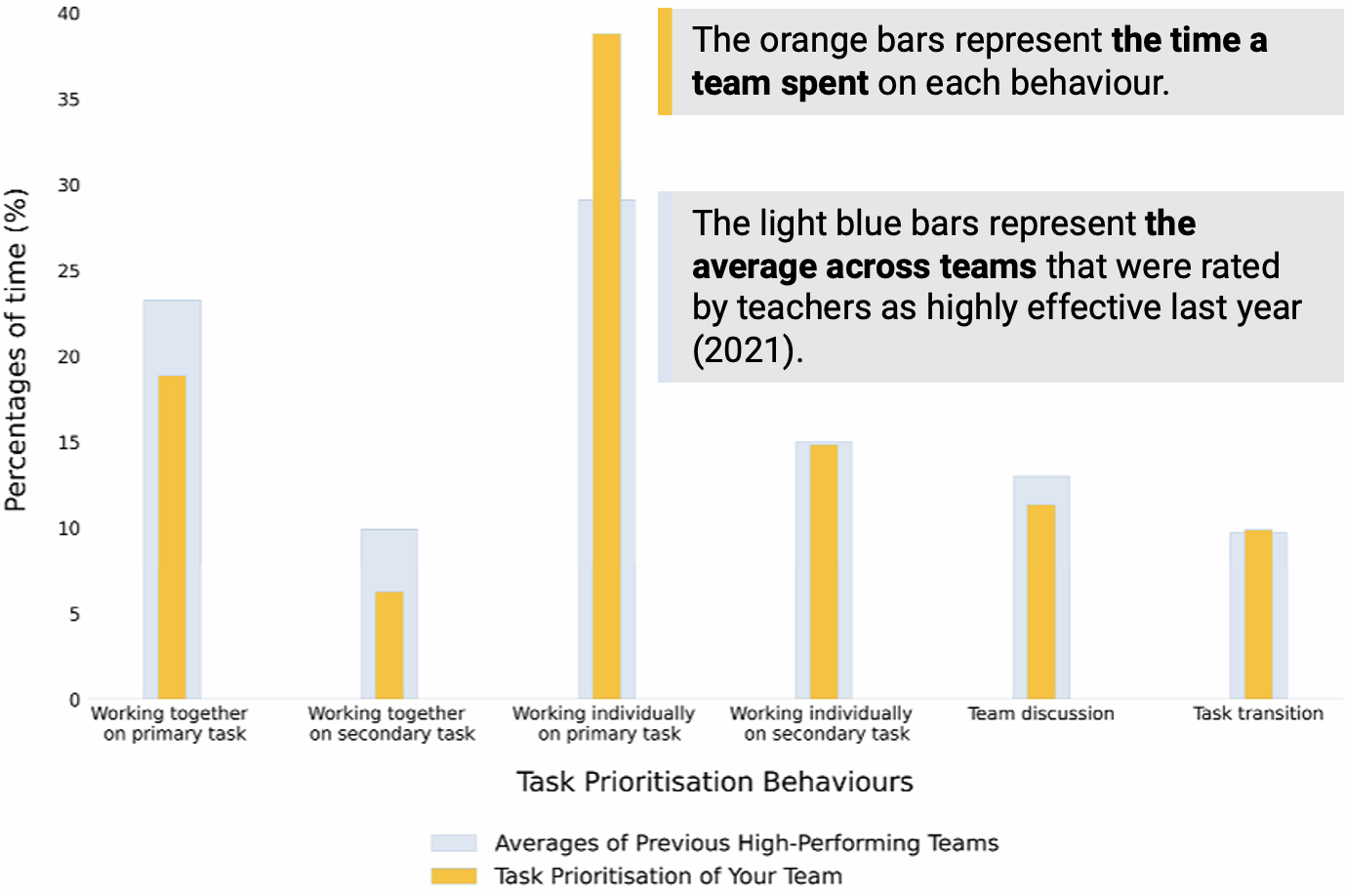}
\end{figure}

\textit{Note: this visualisation IS representing YOUR data}

- To what extent you would trust this visualisation to judge or reflect on your own performance?

        \begin{enumerate}
        \item I’d absolutely trust on it
        
        \item I’d trust it to some extent
        
        \item neutral/borderline
        
        \item I’d not trust it to some extent
        
        \item I’d absolutely not trust on it
        \end{enumerate}
        
- Please, briefly explain your response.

- Do you have any comments on how this visualisation can be improved, or would you add something to the visualisation to make it more straightforward?   
\end{flushleft}

\pagebreak

\textbf{Authors' statement}

The content and contribution of the manuscript are unique in relation to our previous publications. In the manuscript, we cite our own papers where details that are not related to the main contribution of the current manuscript can be found. More specifically, these are two other papers, cited in Section 3, where some details about the human-centred design approach we followed can be consulted:

\textit{ Vanessa Echeverria, Roberto Martinez-Maldonado, Lixiang Yan, Linxuan Zhao, Gloria Fernandez-Nieto, Dragan Gašević, and Simon Buckingham Shum. 2022. HuCETA: A Framework for Human-Centered Embodied Teamwork Analytics. IEEE Pervasive Computing (2022), 1–11.}

\textit{Gloria Milena Fernandez Nieto, Kirsty Kitto, Simon Buckingham Shum, and Roberto Martinez-Maldonado. 2022. Beyond the Learning Analytics Dashboard: Alternative Ways to Communicate Student Data Insights Combining Visualisation, Narrative and Storytelling. In 12th International Learning Analytics and Knowledge Conference (Online, USA) (LAK22). ACM New York, NY, USA, 219–229.}

\end{document}